\documentclass[twocolumn]{aastex631}

\usepackage{footmisc}
\usepackage{amsmath}
\usepackage{longtable}

\shorttitle{The ELM Survey South II}
\shortauthors{Kosakowski et al.}
\graphicspath{{./}{figures/}}

\begin{document}

\title{The ELM Survey South. II.\\Two dozen new low mass white dwarf binaries.}

\author[0000-0002-9878-1647]{Alekzander Kosakowski}
\affiliation{Department of Physics and Astronomy Texas Tech University 
2500 Broadway 
Lubbock, Texas 79409, USA}

\author[0000-0002-4462-2341]{Warren R. Brown}
\affiliation{Center for Astrophysics, Smithsonian Astrophysical Observatory
60 Garden St.,
Cambridge, MA, 012138 USA}

\author[0000-0001-6098-2235]{Mukremin Kilic}
\affiliation{Homer L. Dodge Department of Physics and Astronomy University of Oklahoma
1667 K Street NW, Suite 800 
Norman, OK 73072, USA}

\author[0000-0002-6540-1484]{Thomas Kupfer}
\affiliation{Department of Physics and Astronomy Texas Tech University 
2500 Broadway 
Lubbock, Texas 79409, USA}

\author[0000-0002-2384-1326]{Antoine B\'{e}dard}
\affiliation{Department of Physics, University of Warwick,
Gibbet Hill Road, Coventry CV4 7AL, United Kingdom}

\author[0000-0002-8655-4308]{A. Gianninas}
\affiliation{Physics Department, Trinity College
330 Summit Street
Hartford, CT 06106, USA}

\author[0000-0001-7077-3664]{Marcel A.~Ag\"{u}eros}
\affiliation{Department of Astronomy, Columbia University, 550 West 120th
Street, New York, NY 10027, USA},\affiliation{Laboratoire d’astrophysique
de Bordeaux, Univ. Bordeaux, CNRS, B18N, Allée Geoffroy Saint-Hilaire,
33615 Pessac, France}

\author[0000-0002-6153-9304]{Manuel Barrientos}
\affiliation{Homer L. Dodge Department of Physics and Astronomy University of Oklahoma
1667 K Street NW, Suite 800 
Norman, OK 73072, USA}

\begin{abstract}
We present the results from our ongoing spectroscopic survey targeting low mass white dwarf binaries, focusing on the southern sky. We used a Gaia DR2 and eDR3 based selection and identified 28 new binaries, including 19 new extremely low mass white dwarfs, one short period, likely eclipsing, DABZ, and two potential LISA binaries. We present orbital and atmospheric parameters for each new binary based on our spectroscopic follow-up.

Four of our new binaries show periodic photometric variability in the TESS 2-minute cadence data, including one new eclipsing double-lined spectroscopic binary. Three others show periodic photometric variability in ZTF, including one new eclipsing binary. We provide estimates for the inclinations and scaled component radii for these ZTF variables, based on light curve modeling to our high-speed photometric follow-up observations.

Our observations have increased the sample of ELM Survey binaries identified in the southern sky to 41, an increase of 64\%. Future time domain surveys, such as BlackGEM and the Vera C. Rubin Observatory Legacy Survey of Space and Time, will efficiently identify the photometric variables in the southern sky and significantly increase the population of southern sky low mass white dwarf binaries, leading to a more complete all-sky population of these systems.
\end{abstract}


\section{Introduction} \label{sec:intro}
Extremely Low Mass (ELM; $M\lesssim0.3~{\rm M_\odot}$) white dwarfs are a relatively rare class of He-core white dwarfs which form after early severe mass loss. Because the main sequence lifetime of low mass stars can exceed a Hubble time, the ELM white dwarfs observed today are not expected to have formed through single star evolution. With an exception for the extreme mass loss of high-metallicity stars \citep[see][]{kilic2007}, ELM white dwarfs are expected to form through binary evolution, including one or more episodes of common envelope evolution \citep{li2019}. The result of this binary evolution is a population of evolved compact binaries containing low mass He-core white dwarfs with evolved companions.

These low mass white dwarf binaries are important for studying both binary evolution and formation rates of various exotic systems. \citet{shen2015} and \citet{brown2016b} show that most white dwarf binaries may merge and form other exotic systems such as extreme Helium stars \citep{zhang2014}, accreting AM CVn binaries \citep{kilic2016}, or massive single white dwarfs \citep{kilic2023}. Binaries with well-constrained physical parameters have also been used to place constraints on the efficiency of common envelope ejection \citep[see][]{scherbak2023}.

The ELM Survey \citep{brown2010,kilic2011,brown2012,kilic2012,brown2013,gianninas2015,brown2016a,brown2020,brown2022} is a spectroscopic survey targeting these low-mass white dwarf binaries based on photometry from sky surveys, such as SDSS \citep{abazajian2003} and PanSTARRs \citep{chambers2016}. \citet{kosakowski2020} expanded the ELM Survey into the southern sky using photometry from SkyMapper \citep{onken2019} and VST ATLAS \citep{shanks2015}, and found that a Gaia based selection is efficient for identifying ELM white dwarf binaries. In total, previous ELM Survey studies have constrained the orbits and atmospheric parameters of 120 unique low mass white dwarf binaries \citep{brown2020, kosakowski2020, brown2022}. Similar studies have created catalogs of ELM white dwarf candidates using Gaia DR2 astrometry \citep[see][]{pelisoli2019} and single-epoch spectroscopy from LAMOST DR8 \citep[see][]{wang2022}, many of which still require follow-up observations to confirm their nature.

In this work, we continue the ELM Survey South with our search for low mass white dwarf binaries in the southern sky using a Gaia DR2 and eDR3. While we focused on objects in the southern sky, our Gaia based selection included many Northern sky objects, which we include in this work.

\section{Target Selection}
\begin{figure}
  \center
  \includegraphics[scale=0.45]{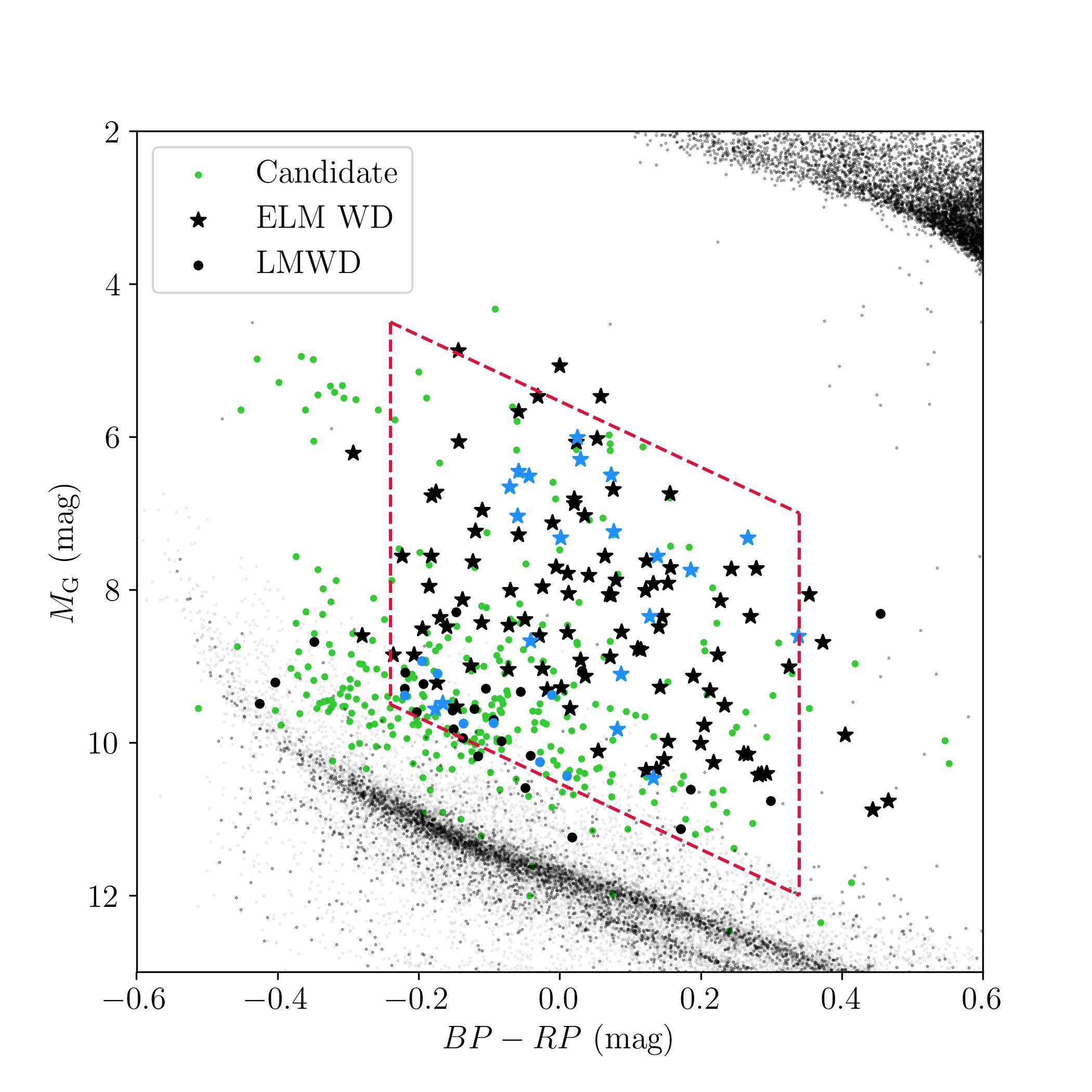}
  \caption{Gaia color-magnitude diagram showing the target selection box described in the text. The 28 new binaries identified in this work are represented with blue symbols, while the previous ELM Survey binaries are represented as black symbols. Stars represent ELM white dwarfs ($M\lesssim0.3~{\rm M_\odot}$). Filled-circles represent low mass white dwarfs. Green circles represent our observed candidates selected through our Gaia DR2 and eDR3 selection.}
  \label{fig:gaia_selection}
\end{figure}

Our target selection made use of parallax and color measurements from Gaia DR2 and eDR3, based on previous ELM Survey discoveries. Figure \ref{fig:gaia_selection} displays the locations of our observed objects on the Gaia DR3 color-magnitude diagram. The 28 new binaries from this work are plotted with blue symbols while previous ELM Survey binaries are plotted as black symbols. Green points represent other objects observed as part of this work with \textsc{parallax\_over\_error>3}, \textsc{parallax>0.5}, and no cuts to \textsc{ruwe}, for which we have obtained at least one optical spectrum. We draw a box around the region surrounding the main concentration of ELM Survey ELM white dwarf binaries, defined by
\begin{align*}
    -0.24<&(BP-RP)<0.34,\\
    M_{\rm G}>4.31&(BP-RP)+5.53,\\
    M_{\rm G}<4.31&(BP-RP)+10.53.
\end{align*}

Within this box, we have observed 217 objects, including the 28 new LMWD binaries presented in this work. Among our observed objects, we identify 27 additional ELM white dwarf candidates based on model atmosphere fits to our follow-up spectroscopy, including 17 which fall within the parameter space of the ``clean" sample of ELM white dwarfs of \cite{brown2020}. Our ongoing follow-up finds that at least half of these spectroscopic candidates show significant radial velocity variability, but have unconstrained orbital parameters. These additional binaries will be presented in a future publication. Most of our observed sample contains single white dwarfs with $\log{g}\approx7.5$. 

\section{Observations} \label{sec:observations}
We used a similar observing strategy as in previous ELM Survey publications: We obtained one optical spectrum for each of our candidates to confirm their nature and perform spectroscopic fitting with model atmospheres. For objects consistent with ELM white dwarfs ($5.0 \lesssim \log{g} \lesssim 7.2$), we obtained multiple additional spectra to check for radial velocity variability and constrain orbital periods. Candidates which do not show significant radial velocity variability, or which show atmospheric parameters inconsistent with ELM white dwarfs, are excluded from extensive follow-up.

Our observing strategy favors identifying short period binaries ($P\lesssim6~{\rm h}$) with large velocity semi-amplitudes, which form through common envelope interaction. Binaries with longer orbital periods and lower velocity semi-amplitudes, which likely form through stable Roche Lobe overflow, are less likely to be detected and constrained through our observing strategy, while those that are identified require more resources to constrain through radial velocity follow-up.

In this section we briefly describe the resources used for our follow-up observations, including the telescopes, instruments, and configurations. 

\subsection{Southern Astrophysical Research Telescope (SOAR)}
We used the SOAR 4.1-meter telescope with the Goodman spectrograph \citep{clemens2004} configured with the $930~{\rm lines}~{\rm mm^{-1}}$ grating and $1.01\arcsec$ slit, resulting in a spectral resolution $\approx2.6~{\rm \AA}$ over the wavelength range $3550-5300~{\rm \AA}$. These data were taken as part of the NOAO programs 2019B-0004, 2020B-0013, and 2021A-0007, and NOIRLab 2022A-161017.

\subsection{Gemini South}
We used the Gemini South 8.1-meter telescope, located on Cerro Pach\'{o}n in Chil\'{e}, with the GMOS-S spectrograph configured with the $0.5\arcsec$ and $1.0\arcsec$ slits and the B600 grating ($600~{\rm lines~mm^{-1}}$) in first-order centered on $\lambda_{center}=5150~{\rm \AA}$. These configurations resulted in resolutions $\approx2.8~{\rm \AA}$ and $\approx5.5~{\rm \AA}$ over the spectral range $3600-6750~{\rm \AA}$, respectively. These data were obtained as part of the programs GN-2021A-Q-300, GS-2020B-Q-304, GS-2021A-Q-300, GS-2021B-Q-304.

\subsection{Walter-Baade Magellan Telescope} 
We used the 6.5-meter Walter-Baade Magellan 1 Telescope at the Las Campanas Observatory in Chil\'{e} with the Magellan Echellette (MagE) spectrograph and the $0.85\arcsec$ slit, resulting in a resolving power $R\approx4800$ covering the wavelength range $3600-7000~{\rm \AA}$.

\subsection{Fred Lawrence Whipple Observatory (FLWO)}
We used the 1.5-meter Tillinghast telescope at FLWO located on Mt. Hopkins in Arizona. Our primary setup used FAST spectrograph with the $1.5\arcsec$ slit and $300~{\rm lines~mm^{-1}}$ grating, resulting in spectral resolution $\approx3.6~{\rm \AA}$ covering the spectral range $3500-7400~{\rm \AA}$. A handful of observations used a slightly different setup with the $1.5~{\rm \AA}$ slit and $600~{\rm lines~mm^{-1}}$ grating, resulting in spectral resolution $\approx1.8~{\rm \AA}$ covering the spectral range $3650-5300~{\rm \AA}$.

\subsection{MMT Observatory}
We used the 6.5-meter MMT with the blue channel spectrograph, primarily with the $1.25\arcsec$ slit and $832~{\rm lines~mm^{-1}}$ grating, resulting in $\approx1.2~{\rm \AA}$ resolution over the wavelength range $3600-4500~{\rm \AA}$. However, a handful of our observations used the $1.0\arcsec$ slit with the $832~{\rm lines~mm^{-1}}$ grating, resulting in spectral resolution $\approx1.0~{\rm \AA}$ over roughly same wavelength range.

\subsection{MDM Observatory}
We used the 2.4-meter Hiltner telescope at the MDM observatory, located in Kitt Peak, Arizona, with the OSMOS spectrograph, the $1.2\arcsec$ slit, and the blue grism ($R\sim1600$), resulting in spectral resolution $3.6~{\rm \AA}$ over the wavelength range $3600-5930~{\rm \AA}$.

\subsection{McDonald Observatory}
We used the 2.1-meter Otto Struve telescope at the McDonald Observatory near Fort Davis, Texas to obtain high-speed photometric follow-up of our binaries to confirm variability seen in various sky survey data archives. We used the ProEM frame-transfer CCD detector with either the BG40, $g$-, $r$-, or $i$-band filters.

\section{Spectroscopic Analysis} \label{sec:analysis}

\subsection{Data Reduction and Calibration}
Data reduction was performed using standard \textsc{iraf} procedures, including bias correction, flat-fielding, aperture extraction, wavelength calibration, and flux calibration with spectro-photometric standard star observations obtained on the same night as each science exposure.

To ensure an accurate wavelength solution for each spectrum, we paired each science exposure with a calibration lamp spectrum taken within $\approx15~{\rm min}$ at the same telescope position as the corresponding science exposure, resulting in wavelength calibration accuracy of $2-3~{\rm km~s^{-1}}$, as tested against night sky lines.

\subsection{Radial Velocities} \label{sec:rv}
We measured the radial velocity of each of our spectra using a cross-correlation method with the \textsc{iraf} package \textsc{xcsao} \citep{kurtz1998}. Each spectrum was cross-correlated with a low-mass white dwarf template spectrum and corrected to zero-velocity. We combined the individual zero-velocity object-specific spectra to create a single high-quality, zero-velocity, spectrum for each object which we use for atmospheric modeling. Finally, we cross-correlated the combined zero-velocity template spectrum for each object with their single-exposure component spectra to obtain our final radial velocity measurements. Our radial velocity measurements for each binary are presented in Table \ref{table:rv_table}. We fit a circular orbit to our radial velocity measurements using a Monte Carlo approach based on \citet{kenyon1986} to estimate orbital period $P$, velocity semi-amplitude $K$, and systemic velocity $\gamma$.

\subsection{Atmospheric Parameters} \label{sec:fits}
We estimated the atmospheric parameters $T_{\rm eff}$ and $\log{g}$ for the primary star in each of our binaries through fitting a grid of 1-D pure-hydrogen atmosphere models to our high signal-to-noise, combined, zero-velocity spectra. The details of this process are described in \citet{gianninas2011,gianninas2014,gianninas2015}. In short: we applied a Levenberg-Marquardt minimization algorithm to fit the normalized Balmer line profiles of ${\rm H}\beta$ through ${\rm H}12$, where visible, to a grid of pure-hydrogen model atmospheres convolved to the spectral resolution of the observed spectra, defined by the observation instrument setup. Our parameter estimates are reliable for binaries in which the companion does not contribute a significant amount to the total system light.

Cool objects, with temperatures $T_{\rm eff}\lesssim10{\rm ,}000~{\rm K}$ return systematically large $\log{g}$ when fit with 1D stellar models \citep[see][]{tremblay2011}. Thus, we apply a 3D correction to the fits of cool objects using the equations provided in \citet{tremblay2015}.

Our minimization process returns internal uncertainties which are most sensitive to the flux calibration and signal-to-noise ratio of the input spectrum. We add in quadrature the external uncertainties of $\sigma_{T_{\rm eff}}=1.4\%$ and $\sigma_{\log{g}}=0.042~{\rm dex}$, as presented in \citet{liebert2005}, to each of our reported values.

Table \ref{table:fig1_table} presents the atmospheric parameters for our observed sample (Figure \ref{fig:gaia_selection}: green points) with $\log{g}>5.0$. A representative optical spectrum for many of these objects is available in an online Zenodo archive\footnote{https://doi.org/10.5281/zenodo.7849976} \citep{zenodo} in \textsc{fits} format.

\section{Archival Light Curve Data}
Large scale time domain surveys are a valuable resource for efficiently identifying transients and periodic variables. While these time domain surveys do not replace conventional target selection methods, when paired with color surveys and the precise astrometry from Gaia, it is possible to efficiently perform follow-up observations for characterizing photometrically variable sources.

We made use of the publicly-available online data archives of the Transiting Exoplanet Survey Satellite \citep[TESS;][]{ricker2015} and the Zwicky Transient Facility \citep[ZTF;][]{bellm2019,graham2019,masci2019} DR10 to confirm photometric variability in our targets and constrain orbital periods.

We used the computational resources of the Texas Tech University High Performance Computing Center to perform searches for periodic variability in each of our target light curves from both the TESS and ZTF public data archives. We used two algorithms to identify different types of photometric variability:

For sinusoidal variability, typically caused by a tidally-distorted star in a compact binary, relativistic beaming, strong reflection effects, stellar rotation, or low-amplitude pulsations, we used the \textsc{astropy} \citep{astropy2013, astropy2018, astropy2022} implementation of the Lomb-Scargle (LS) periodogram \citep{lomb1976,scargle1982,vanderplas2018}, searching periods between $5~{\rm min}$ and $684~{\rm min}$.

To identify eclipsing binaries, we make use of a generic Box Least Squares \citep[BLS;][]{kovacs2002} algorithm, which attempts to fit box-shaped eclipses to light curve data phase-folded at periods within the provided frequency grid. Specifically, our BLS algorithm searched for eclipse duration between 0.1\% and 10.0\% of the orbit and orbital periods between $5~{\rm min}$ and $684~{\rm min}$. The BLS algorithm is ideal for identifying eclipsing systems with sharp ingress and egress features, such as eclipsing binaries containing white dwarfs.

\subsection{TESS High Cadence Data Archive}
TESS is a space-based all-sky survey satellite designed to identify exoplanets through transit detections in 27-day long pointings, covering a $24\times96~{\rm deg^2}$ field of view in a broadband filter ($\approx6000-10{\rm ,}000~{\rm \AA}$). The original mission (2018-2020) obtained 30-minute cadence data in Full Frame Images (FFI) with select fields obtaining 2-minute high-cadence observations. The recent extended mission (2020-2022) has improved these to 10-minute cadence FFIs and 20-second high-cadence fields. While the plate scale for TESS is large ($\approx21\arcsec~{\rm px^{-1}}$), objects that show photometric variability in TESS data are typically very well-sampled.

We searched the Barbara A. Mikulski Archive for Space Telescopes (MAST), through the Python module \textsc{astroquery.mast}\footnote{https://astroquery.readthedocs.io/en/latest/mast/mast.html}, to identify the TESS Input Catalog \citep[TIC;][]{stassun2018} target ID for each of our objects. Using the TIC ID, we downloaded the TESS 2-minute and 20-second cadence light curve data through the online TESS archive\footnote{https://archive.stsci.edu/tess/bulk\_downloads/bulk\_downloads\_ffi-tp-lc-dv.html}.

For each sector of TESS data we recovered, we removed lower quality data, based on bit flags described on the TESS Data Quality Overview webpage\footnote{https://outerspace.stsci.edu/display/TESS/2.0+-+Data+Product+Overview}, following the steps outlined in the Jupyter Notebook examples provided through the Space Telescope Science Institute's GitHub page\footnote{https://github.com/spacetelescope/notebooks/}. We combined data across multiple sectors by simply dividing each sector's light curve by its median PDCSAP flux value (aperture photometry corrected for common instrumental systematics and trends) and appending each scaled sector light curve together.

\subsection{ZTF Data Archive}
ZTF is an optical time-domain survey designed to image the entire Northern sky down to $\approx20.5~{\rm mag}$ every two days in two filters, ${\rm ZTF}-g$ and ${\rm ZTF}-r$, with ${\rm ZTF}-i$ sampled less frequently. The ZTF survey uses the 48-inch Schmidt Telescope at the Palomar Observatory in California with a $48~{\rm deg^2}$ field of view.

We performed a cone-search on the public Data Release\footnote{https://www.ztf.caltech.edu/ztf-public-releases.html} 10 (DR10) data archive using a $5\arcsec$ search radius centered on the Gaia DR2 or eDR3 coordinates for each of our objects. Because ZTF assigns different object IDs for the same object in different filters, we combined data for object detections within $2.5\arcsec$ of each coordinate pair returned within our $5\arcsec$ search radius. This process separates nearby ($2.5\arcsec<d<5.0\arcsec$) objects in relatively crowded fields.

To increase the temporal sampling of the ZTF light curves, we artificially shifted the $r$-band and $i$-band data such that their median magnitudes matched the median value of the $g$-band data. We then used this median-combined light curve in our periodicity search.

Many of our objects which show short-period photometric variability in ZTF were included in the ZTF deep-drilling survey, which targets specific fields for continuous observations over $\approx90~{\rm min}$ in the ZTF-$r$ band \citep{kupfer2021}. These deep drilling fields provide significant orbital phase coverage that would otherwise be relatively sparse in the standard ZTF northern sky survey.

\section{Results} \label{sec:results}

\begin{figure}
  \includegraphics[scale=0.5]{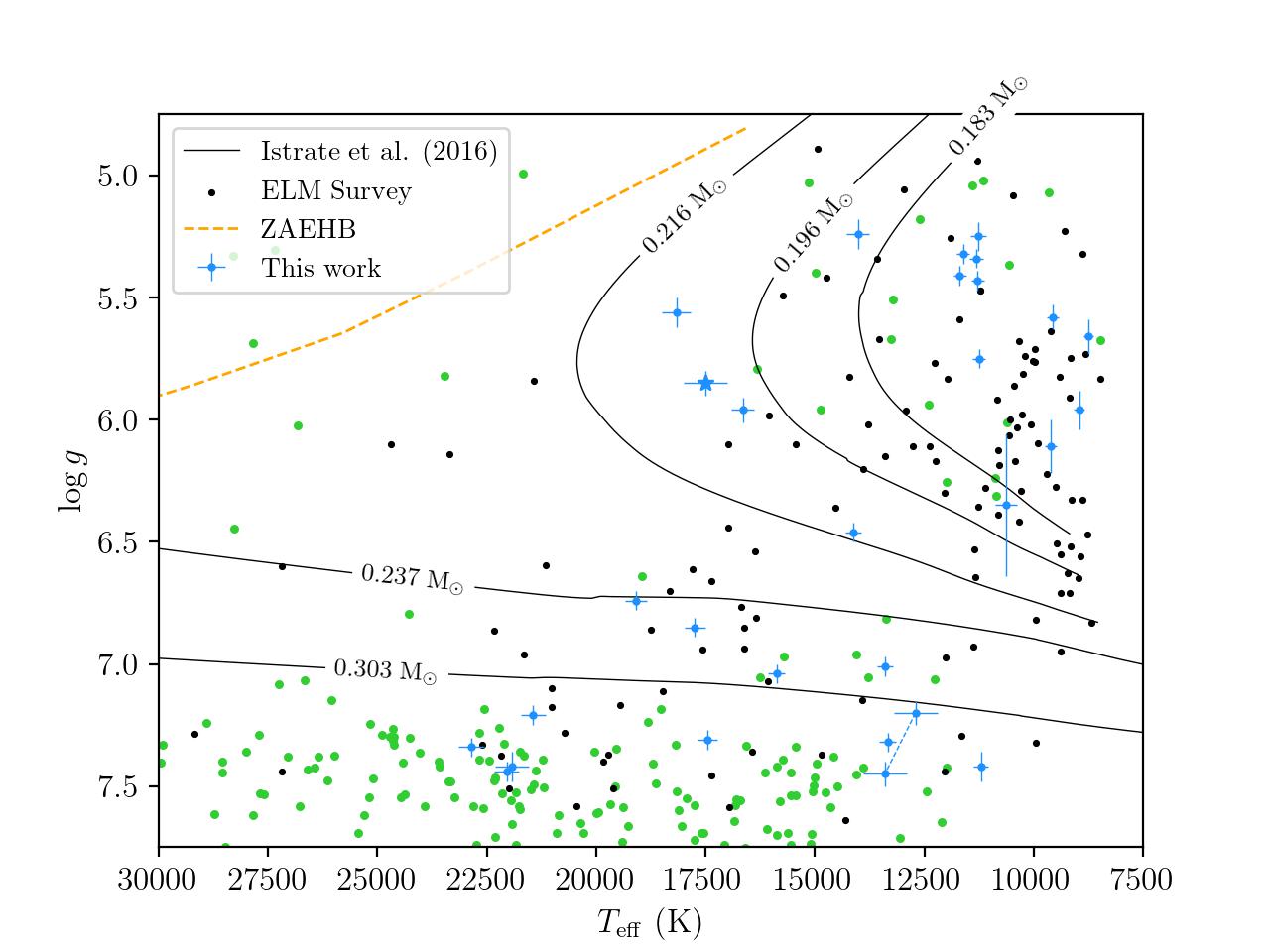}
  \caption{$\log{g}$-$T_{\rm eff}$ distribution of low mass white dwarfs identified as part of the ongoing ELM Survey. Previous ELM Survey binaries are displayed as black points. New binaries from this work are colored blue. Other objects observed as part of this work are colored green. We overplot the cooling tracks for $0.183~{\rm M_\odot}$, $0.196~{\rm M_\odot}$, $0.216~{\rm M_\odot}$, $0.237~{\rm M_\odot}$, and $0.303~{\rm M_\odot}$ white dwarfs with $Z=0.001$ from \citet{istrate2016}, including rotation and diffusion. We mark the location of the DABZ J2049$+$3351 with a blue star symbol and use a dashed line to connect the markers for the individual stars in the double-lined binary J2102$-$4145. The Zero-Age Extreme Horizontal Branch (ZAEHB) is colored as an orange dashed line for reference.}
  \label{fig:teff_logg_diagram}
\end{figure}

We have constrained the atmospheric parameters and orbital periods for 28 new binaries identified through our target selections. In this section, we present details for the eight binaries which require additional explanation, such as those with additional constraints from light curve data or those with unusual spectra. The remaining objects are summarized in Table \ref{table:spectroscopy_table}. We display the $\log{g}$-$T_{\rm eff}$ distribution for each of these objects in Figure \ref{fig:teff_logg_diagram}.

\begin{table*}
\center
  \renewcommand{\arraystretch}{1.15} 
  \addtolength{\tabcolsep}{2pt}
	\begin{tabular}{l c c c c c c c}
    \hline
    \hline
    \multicolumn{1}{C}{\rm Object~Name} &
    \multicolumn{1}{C}{\rm R.A.} & \multicolumn{1}{C}{\rm Decl.} &
    \multicolumn{1}{C}{$T_{\rm eff}$} &
    \multicolumn{1}{C}{$\log{g}$} &
    \multicolumn{1}{C}{$M_1$} &
    \multicolumn{1}{C}{${\rm Gaia}~G$} &
    \multicolumn{1}{C}{${\rm Gaia~Parallax}$} \\
    \multicolumn{1}{C}{} &
    \multicolumn{1}{C}{(2016.0)} & \multicolumn{1}{C}{(2016.0)} &
    \multicolumn{1}{C}{($K$)} &
    \multicolumn{1}{C}{(${\rm cm~s^{-2}}$)} &
    \multicolumn{1}{C}{($M_\odot$)} &
    \multicolumn{1}{C}{(${\rm mag}$)} &
    \multicolumn{1}{C}{(${\rm mas}$)} \\
    \hline
    {J0135$+$2359} & {01:35:00.856} & {$+$23:59:46.091} & {$14130\pm210$} & {$6.46\pm0.05$} & {$0.21\pm0.04$} & {$18.7$} & {$1.18\pm0.29$} \\
    {J0155$-$4148} & {01:55:34.866} & {$-$41:48:18.433} & {$11250\pm170$} & {$5.75\pm0.05$} & {$0.22\pm0.02$} & {$15.7$} & {$2.08\pm0.03$} \\
    {J0215$+$0155}\footnote{Photometric variability: TESS high-cadence\label{footnote:tess}}\footref{footnote:pelisoli}  & {02:15:06.244} & {$+$01:55:03.363} & {$11310\pm180$} & {$5.34\pm0.05$} & {$0.29\pm0.02$} & {$14.3$} & {$2.15\pm0.03$} \\
    {J0221$+$1710}\footnote{Photometric variability: ZTF\label{footnote:ztf}}\footref{footnote:pelisoli} & {02:21:10.832} & {$+$17:10:49.182} & {$13400\pm200$} & {$7.01\pm0.04$} & {$0.27\pm0.01$} & {$17.7$} & {$3.58\pm0.12$} \\
    {J0256$+$4405}\footnote{\citet{pelisoli2019} ELM white dwarf candidate\label{footnote:pelisoli}} & {02:56:35.153} & {$+$44:05:27.363} & {$18170\pm350$} & {$5.56\pm0.06$} & {$0.22\pm0.02$} & {$15.8$} & {$1.40\pm0.04$} \\
    {J0450$-$0145}\footref{footnote:pelisoli} & {04:50:13.108} & {$-$01:45:48.150} & {$9560\pm140$} & {$5.58\pm0.06$} & {$0.19\pm0.02$} & {$17.7$} & {$0.91\pm0.11$} \\
    {J0501$-$2312}\footref{footnote:pelisoli} & {05:01:29.865} & {$-$23:12:04.397} & {$21440\pm330$} & {$7.21\pm0.05$} & {$0.36\pm0.01$} & {$18.0$} & {$1.64\pm0.10$} \\
    {J0517$-$1153}\footref{footnote:pelisoli} & {05:17:24.974} & {$-$11:53:25.849} & {$16650\pm300$} & {$5.96\pm0.06$} & {$0.19\pm0.02$} & {$16.2$} & {$1.47\pm0.04$} \\
    {J0545$-$1902} & {05:45:45.301} & {$-$19:02:45.499} & {$22850\pm340$} & {$7.34\pm0.05$} & {$0.40\pm0.02$} & {$17.3$} & {$2.59\pm0.07$} \\
    {J0725$-$1245} & {07:25:27.362} & {$-$12:45:46.824} & {$21920\pm420$} & {$7.42\pm0.06$} & {$0.42\pm0.02$} & {$18.9$} & {$1.51\pm0.25$} \\
    {J1121$+$6052}\footref{footnote:tess} & {11:21:57.163} & {$+$60:52:10.265} & {$11690\pm170$} & {$5.41\pm0.05$} & {$0.19\pm0.01$} & {$16.0$} & {$1.33\pm0.04$} \\
    {J1129$+$4715}\footref{footnote:pelisoli} & {11:29:14.162} & {$+$47:15:01.726} & {$11610\pm170$} & {$5.32\pm0.05$} & {$0.19\pm0.01$} & {$16.1$} & {$1.18\pm0.04$} \\
    {J1240$-$0958} & {12:40:32.501} & {$-$09:58:59.603} & {$14020\pm280$} & {$5.24\pm0.06$} & {$0.20\pm0.02$} & {$19.0$} & {$1.30\pm0.30$} \\
    {J1255$-$1853} & {12:55:39.147} & {$-$18:53:32.101} & {$11270\pm200$} & {$5.25\pm0.06$} & {$0.19\pm0.01$} & {$17.8$} & {$0.55\pm0.13$} \\
    {J1459$-$1920} & {14:59:02.159} & {$-$19:20:33.552} & {$8740\pm130$} & {$5.66\pm0.07$} & {$0.26\pm0.02$} & {$18.1$} & {$0.71\pm0.16$} \\
    {J1506$-$1125} & {15:06:12.345} & {$-$11:25:11.994} & {($22050\pm320$)} & {($7.44\pm0.05$)} & {($0.43\pm0.02$)} & {$17.0$} & {$2.42\pm0.10$} \\
    {J1526$-$2711}\footref{footnote:pelisoli} & {15:26:01.115} & {$-$27:11:56.660} & {$17460\pm260$} & {$7.31\pm0.05$} & {$0.37\pm0.02$} & {$18.3$} & {$1.61\pm0.18$} \\
    {J1553$+$6736}\footref{footnote:pelisoli} & {15:53:28.008} & {$+$67:36:10.560} & {$9610\pm150$} & {$6.11\pm0.11$} & {$0.22\pm0.04$} & {$16.5$} & {$2.36\pm0.04$} \\
    {J1555$+$1007}\footref{footnote:pelisoli} & {15:55:15.894} & {$+$10:07:24.851} & {$13340\pm220$} & {$7.32\pm0.05$} & {$0.35\pm0.02$} & {$18.2$} & {$2.52\pm0.15$} \\
    {J1657$-$0417} & {16:57:24.888} & {$-$04:17:22.348} & {$17750\pm270$} & {$6.85\pm0.05$} & {$0.27\pm0.02$} & {$18.3$} & {$2.04\pm0.18$} \\
    {J1808$+$2723} & {18:08:38.994} & {$+$27:23:12.216} & {$10630\pm270$} & {$6.35\pm0.29$} & {$0.22\pm0.04$} & {$15.5$} & {$2.82\pm0.03$} \\
    {J1812$+$0525}\footref{footnote:ztf} & {18:12:38.471} & {$+$05:25:29.868} & {$8960\pm130$} & {$5.96\pm0.08$} & {$0.28\pm0.03$} & {$18.9$} & {$0.85\pm0.26$} \\
    {J1832$+$2031}\footref{footnote:pelisoli} & {18:32:36.539} & {$+$20:31:08.202} & {$19080\pm290$} & {$6.74\pm0.05$} & {$0.29\pm0.03$} & {$17.6$} & {$1.61\pm0.08$} \\
    {J2013$-$1310}\footref{footnote:pelisoli} & {20:13:53.498} & {$-$13:10:41.750} & {$11200\pm190$} & {$7.42\pm0.06$} & {$0.37\pm0.02$} & {$18.7$} & {$2.21\pm0.25$} \\
    {J2049$+$3351}\footref{footnote:ztf} & {20:49:51.274} & {$+$33:51:53.126} & {$17500\pm500$} & {$5.85\pm0.05$} & {\nodata} & {$18.7$} & {$0.51\pm0.16$} \\
    {J2102$-$4145a}\footref{footnote:tess} & {21:02:20.456} & {$-$41:45:01.736} & $12700\pm500$ & $7.20\pm0.05$ & $0.32\pm0.01$ & {$15.8$} & {$6.07\pm0.04$} \\
    {J2102$-$4145b}\footref{footnote:tess} & {21:02:20.456} & {$-$41:45:01.736} & $13400\pm500$ & $7.45\pm0.05$ & $0.39\pm0.01$ & {\nodata} & {$6.07\pm0.04$} \\
    {J2243$-$4511} & {22:43:27.479} & {$-$45:11:18.404} & {$15880\pm230$} & {$7.04\pm0.05$} & {$0.29\pm0.01$} & {$17.4$} & {$2.57\pm0.09$} \\
    {J2303$-$2614}\footref{footnote:tess}\footref{footnote:pelisoli} & {23:03:23.542} & {$-$26:14:59.917} & {$11280\pm160$} & {$5.43\pm0.05$} & {$0.18\pm0.01$} & {$13.8$} & {$3.12\pm0.02$} \\
    \hline
    \hline
	\end{tabular}
    \caption{White dwarf parameters determined through optical spectroscopy for the 28 new binaries identified in this work. Our reported atmospheric parameters $T_{\rm eff}$ and $\log{g}$ include the external uncertainties of $\sigma_{T_{\rm eff}}=1.4\%$ and $\sigma_{\log{g}}=0.042~{\rm dex}$ from \citet{liebert2005}. Binaries which show photometric variability in the TESS 2$-$minute cadence or ZTF data archives are marked. We apply 3D corrections using the equations from \citet{tremblay2015} for objects cooler than $T_{\rm eff}\approx10000~{\rm K}$. The atmospheric parameter values for J1506$-$1125 displayed in this table are based on single-star models. We describe our multi-component modeling to J1506$-$1125 in Section \ref{sec:results}, which does not identify a unique solution.}
    \label{table:spectroscopy_table}
\end{table*}
\begin{table*}
\center
  \renewcommand{\arraystretch}{1.15} 
  \addtolength{\tabcolsep}{2pt}
	\begin{tabular}{l c r r c c c}
    \hline
    \hline
    \multicolumn{1}{C}{\rm Object~Name} &
    \multicolumn{1}{C}{$P$} &
    \multicolumn{1}{C}{$K$} &
    \multicolumn{1}{C}{$\gamma$} &
    \multicolumn{1}{C}{$M_{\rm 2}$} &
    \multicolumn{1}{C}{$\tau_{\rm merge}$} &
    \multicolumn{1}{C}{\rm Disk} \\
    \multicolumn{1}{C}{} &
    \multicolumn{1}{C}{(${\rm days}$)} &
    \multicolumn{1}{C}{(${\rm km~s^{-1}}$)} &
    \multicolumn{1}{C}{(${\rm km~s^{-1}}$)} &
    \multicolumn{1}{C}{(\rm $M_\odot$)} &
    \multicolumn{1}{C}{(${\rm Gyr}$)} &
    \multicolumn{1}{C}{} \\
    \hline
    J0135$+$2359 & $1.177655\pm0.009923$ & $178.9\pm6.4$ & $-35.7\pm 6.2$ & $>1.02\pm0.09$ & $<371.4\pm71.7$ & {1} \\
    J0155$-$4148 & $0.343865\pm0.000317$ & $220.4\pm3.7$ & $-4.0\pm 2.9$ & $>0.67\pm0.03$ & $<18.1\pm1.6$ & {1} \\
    J0215$+$0155\footnote{Photometric variability: TESS high-cadence\label{footnote:tess}}\footref{footnote:pelisoli} & $0.387941\pm0.000001$ & $186.4\pm  1.5$ & $-49.4\pm 1.1$ & $>0.58\pm0.02$ & $<21.7\pm1.4$ & {0} \\
    J0221$+$1710\footnote{Photometric variability: ZTF\label{footnote:ztf}}\footref{footnote:pelisoli} & $0.061288\pm0.000020$ & $347.9\pm  4.2$ & $35.3\pm 3.7$ & $0.58\pm0.02$ & $0.17\pm0.01$ & {1} \\
    J0256$+$4405\footnote{\citet{pelisoli2019} ELM white dwarf candidate\label{footnote:pelisoli}} & $0.261260\pm0.000087$ & $243.7\pm  3.8$ & $28.1\pm 2.9$ & $>0.68\pm0.03$ & $<8.6\pm0.8$ & {1} \\
    J0450$-$0145\footref{footnote:pelisoli} & $0.192169\pm0.000040$ & $260.2\pm  3.3$ & $65.5\pm 2.9$ & $>0.61\pm0.02$ & $<4.7\pm0.5$ & {1} \\
    J0501$-$2312\footref{footnote:pelisoli} & $0.086593\pm0.001156$ & $105.1\pm  5.1$ & $ 6.3\pm 7.2$ & $>0.14\pm0.01$ & $<1.1\pm0.1$ & {1} \\
    J0517$-$1153\footref{footnote:pelisoli} & $0.250521\pm0.000001$ & $309.7\pm  3.1$ & $53.5\pm 2.6$ & $>1.07\pm0.04$ & $<6.3\pm0.7$ & {1} \\
    J0545$-$1902 & $0.144472\pm0.000684$ & $134.7\pm  5.4$ & $146.1\pm 5.7$ & $>0.25\pm0.02$ & $<2.4\pm0.2$ & {1} \\
    J0725$-$1245 & $0.106135\pm0.000061$ & $ 79.6\pm  5.0$ & $90.4\pm 3.2$ & $>0.12\pm0.01$ & $<2.0\pm0.2$ & {1} \\
    J1121$+$6052\footref{footnote:tess} & $0.084511\pm0.000013$ & $183.5\pm  2.6$ & $-15.7\pm 2.2$ & $>0.20\pm0.01$ & $<1.3\pm0.1$ & {1} \\
    J1129$+$4715\footref{footnote:pelisoli} & $0.238823\pm0.000032$ & $185.8\pm  4.4$ & $40.9\pm 3.0$ & $>0.37\pm0.02$ & $<12.4\pm0.7$ & {1} \\
    J1240$-$0958 & $0.400383\pm0.002945$ & $209.8\pm  6.1$ & $23.9\pm 3.3$ & $>0.65\pm0.04$ & $<30.3\pm3.2$ & {1} \\
    J1255$-$1853 & $0.363739\pm0.001501$ & $230.8\pm  6.2$ & $-15.2\pm29.5$ & $>0.73\pm0.04$ & $<22.6\pm1.5$ & {1} \\
    J1459$-$1920 & $0.151990\pm0.000030$ & $287.8\pm  7.4$ & $45.6\pm 5.0$ & $>0.70\pm0.04$ & $<1.7\pm0.1$ & {0} \\
    J1506$-$1125 & $0.032320\pm0.000390$ & $167.5\pm  4.3$ & $43.5\pm 2.8$ & $>0.18\pm0.01$ & $<0.056\pm0.003$ & {1} \\
    J1526$-$2711\footref{footnote:pelisoli} & $0.027982\pm0.000439$ & $336.0\pm  5.6$ & $ 8.4\pm 4.8$ & $>0.40\pm0.02$ & $<0.021\pm0.001$ & {1} \\
    J1553$+$6736\footref{footnote:pelisoli} & $0.174522\pm0.000431$ & $ 91.6\pm  5.4$ & $10.9\pm 6.1$ & $>0.12\pm0.01$ & $<12.4\pm2.3$ & {1} \\
    J1555$+$1007\footref{footnote:pelisoli} & $0.298037\pm0.000877$ & $148.5\pm  6.7$ & $-51.4\pm 3.7$ & $>0.38\pm0.03$ & $<13.0\pm1.0$ & {1} \\
    J1657$-$0417 & $0.083954\pm0.000441$ & $289.4\pm  8.8$ & $-69.8\pm21.3$ & $>0.50\pm0.03$ & $<0.44\pm0.04$ & {1} \\
    J1808$+$2723 & $0.098787\pm0.000053$ & $187.2\pm  3.0$ & $-69.3\pm 1.9$ & $>0.24\pm0.02$ & $<1.4\pm0.2$ & {1} \\
    J1812$+$0525\footref{footnote:ztf} & $0.059847\pm0.000083$ & $373.3\pm  6.2$ & $-139.6\pm 4.7$ & $0.73^{+0.05}_{-0.04}$ & $0.13\pm  0.01$ & {0} \\
    J1832$+$2031\footref{footnote:pelisoli} & $0.046641\pm0.000002$ & $335.2\pm  4.2$ & $-37.4\pm 3.3$ & $>0.47\pm0.02$ & $<0.090\pm0.009$ & {1} \\
    J2013$-$1310\footref{footnote:pelisoli} & $0.061618\pm0.000597$ & $300.9\pm  6.5$ & $-40.2\pm 6.3$ & $>0.51\pm0.02$ & $<0.14\pm0.01$ & {1} \\
    J2049$+$3351\footref{footnote:ztf} & $0.029747\pm0.000007$ & $513.2\pm  9.5$ & $-3.4\pm 7.7$ & \nodata & \nodata & {1} \\
    J2102$-$4145a\footref{footnote:tess} & $0.117631\pm0.003244$ & $227^{+8}_{-6}$ & $-7.9^{+7.1}_{-5.8}$ & \nodata & $<0.74\pm0.02$ & {1} \\
    J2102$-$4145b\footref{footnote:tess} & $0.117631\pm0.003244$ & $186^{+8}_{-7}$ & $-18.7^{+6.2}_{-7.0}$ & \nodata & $<0.74\pm0.02$ & {1} \\
    J2243$-$4511 & $0.109479\pm0.000043$ & $249.4\pm  4.9$ & $ 6.0\pm 3.7$ & $>0.46\pm0.02$ & $<0.89\pm0.04$ & {1} \\
    J2303$-$2614\footref{footnote:tess}\footref{footnote:pelisoli} & $0.118195\pm0.000032$ & $302.9\pm  2.3$ & $-17.1\pm 2.1$ & $>0.58\pm0.01$ & $<1.4\pm0.1$ & {1} \\
    \hline
    \hline
	\end{tabular}
    \caption{Orbital solutions to the 28 new low mass white dwarf binaries presented in this work.}
    \label{table:orbit_table}
\end{table*}

\newpage 
\subsection*{\rm J0215$+$0155}
\begin{figure*}
  \center
  \includegraphics[scale=0.3]{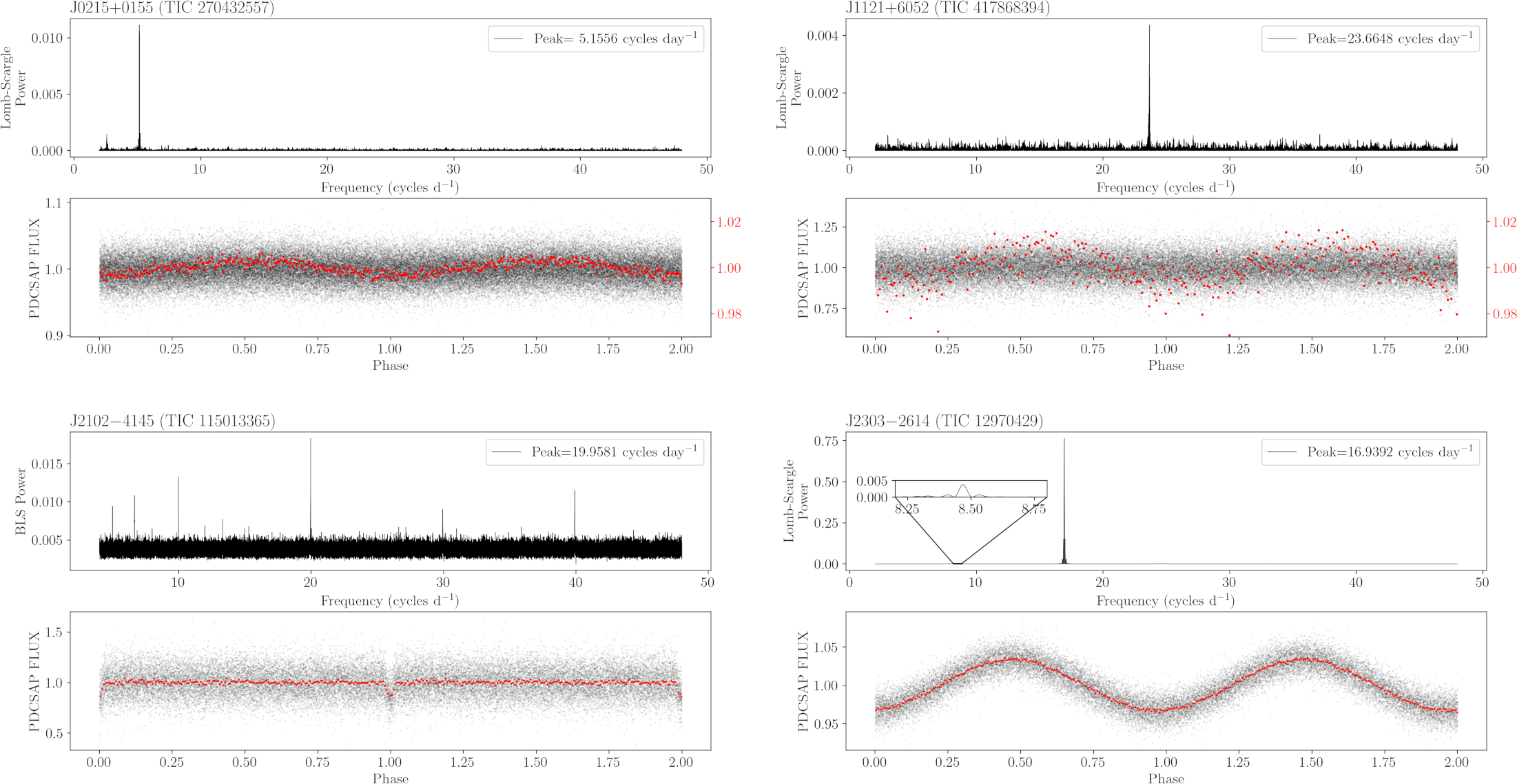}
  \caption{Lomb-Scargle or Box Least Squares (BLS) power spectrum (top) and its TESS 2-minute cadence light curve (bottom) for J0215$+$0155 (TIC 270432557; upper-left), J1121$+$6052 (TIC 417868394; upper-right), J2102$-$4145 (TIC 115013365; lower-left), and J2303$-$2614 (TIC 12970429; lower-right).
  Red data points represent the original data binned by 100. We re-scaled the binned data of J0215$+$0155 and J1121$+$6052 to emphasize the variability seen at the 1\% level.}
  \label{fig:combined_tess}
\end{figure*}

We obtained 112 radial velocity measurements for J021506.244$+$015503.363 (J0215$+$0155; GAIA DR3 2513538251735261696), resulting in best-fitting orbital parameters $P_{\rm RV}=9.3106~{\rm h}$, $K=186.4\pm1.5~{\rm km~s^{-1}}$, and $\gamma=-49.4\pm1.1~{\rm km~s^{-1}}$. Together with Gaia DR3 astrometry, we estimated the Galactic space velocities $[U,V,W]=[121\pm1,-37\pm1,-59\pm1]~{\rm km~s^{-1}}$,  ($U$ positive towards the Galactic center), corrected for the motion of the local standard of rest \citep{schonrich2010}.

We determine Galactic disk and halo membership by computing the Mahalanobis distance between the measured Galactic space velocities and the velocity distributions for thick disk and halo populations using average velocities and velocity dispersions from \citep{chiba2000}. Specifically, we compared against velocity distributions defined by $[\langle U \rangle,\langle V \rangle,\langle W \rangle]_{\rm Disk}=[4\pm46,-20\pm50,-3\pm35]~{\rm km~s^{-1}}$ and $[\langle U \rangle,\langle V \rangle,\langle W \rangle]_{\rm Halo}=[17\pm141,-187\pm106,-5\pm94]~{\rm km~s^{-1}}$ for the thick disk and halo, respectively. Our measurements for J0215$+$0155 are consistent with Galactic halo membership.

Our pure-hydrogen model atmosphere fits to the summed zero-velocity spectrum of J0215$+$0155 result in best-fitting atmospheric parameters $T_{\rm eff}=11{\rm ,}310\pm180~{\rm K}$ and $\log{g}=5.34\pm0.05$, corresponding to white dwarf with mass $M_1=0.29\pm0.02~{\rm M_{\odot}}$ based on the halo metallicity models for He-core white dwarfs from \citet{istrate2016}. With the velocity semi-amplitude known and orbital period known, we used the binary mass function,
\begin{equation}
    \frac{(M_2\sin{i})^3}{(M_1+M_2)^2} = \frac{PK^3}{2\pi G}
\end{equation}
to estimate the minimum companion mass $M_{\rm 2,min}=0.59\pm0.02~{\rm M_\odot}$.

We find three sectors of TESS 2-minute cadence data for J0215$+$0155 (TIC 270432557). Our Lomb Scargle algorithm identifies weak ($\sim1\%$ level) periodic variability at $f_{\rm TESS}=5.1556~{\rm cycles~d^{-1}}$, equal to half of the orbital period determined through radial velocity measurements. Figure \ref{fig:combined_tess} (upper-left) displays the phase-folded TESS light curve of J0215$+$0155 and its Lomb Scargle power spectrum. A smaller peak can be seen in the power spectrum at the true orbital frequency of the system. The dominant frequency at $5.1556~{\rm cycles~d^{-1}}$ is likely caused by tidal distortions in the compact binary. Detailed high-precision follow-up light curve analysis may help place constraints on the orbital inclination and mass ratio of this binary, \citep[see, for example,][]{hermes2014}.

\newpage
\subsection*{\rm J0221$+$1710}
We obtained 19 radial velocity measurements of J022110.832$+$171049.182 (J0221$+$1710; GAIA DR3 79808158877017216), resulting in orbital parameters $P_{\rm RV}=1.4709\pm0.0005~{\rm h}$, $K=347.9\pm4.2~{\rm km~s^{-1}}$, $\gamma=35.3\pm3.7~{\rm km~s^{-1}}$. With precise Gaia astrometry, we estimated Galactic space velocities $[U,V,W]=[-40.4\pm1.8, -4.8\pm1.6, 7.5\pm1.7]~{\rm km~s^{-1}}$, which are well within the $2\sigma$ velocity ellipsoid for the Galactic disk.

Our best-fitting pure-hydrogen atmospheric parameters to J0221$+$1710 are $T_{\rm eff}=13{\rm ,}400\pm200~{\rm K}$ and $\log{g}=7.01\pm0.04$, corresponding to a white dwarf with mass $M_1=0.27\pm0.01~{\rm M_{\odot}}$, based on the ${\rm Z}=0.02$ model tracks of \citet{istrate2016}.

Our BLS periodicity search on the ZTF DR13 data archive identified periodic $\Delta m\approx0.2~{\rm mag}$ eclipses at $P_{\rm ZTF,BLS}=88.2508~{\rm min}$ ($1.4708~{\rm h}$), in agreement with our orbital period obtained through radial velocity measurements.

We obtained high-speed $g$- and $r$-band photometry using the McDonald 2.1-meter telescope on UT 20221001 and UT 20221002. Our observations cover three eclipses in each filter. We used ${\rm EXPTIME}=10~{\rm s}$, resulting in five mid-eclipse data points and four data points during the ingress or egress per eclipse for the $\approx90~{\rm s}$ eclipse duration.

\begin{figure*}
  \includegraphics[scale=0.35]{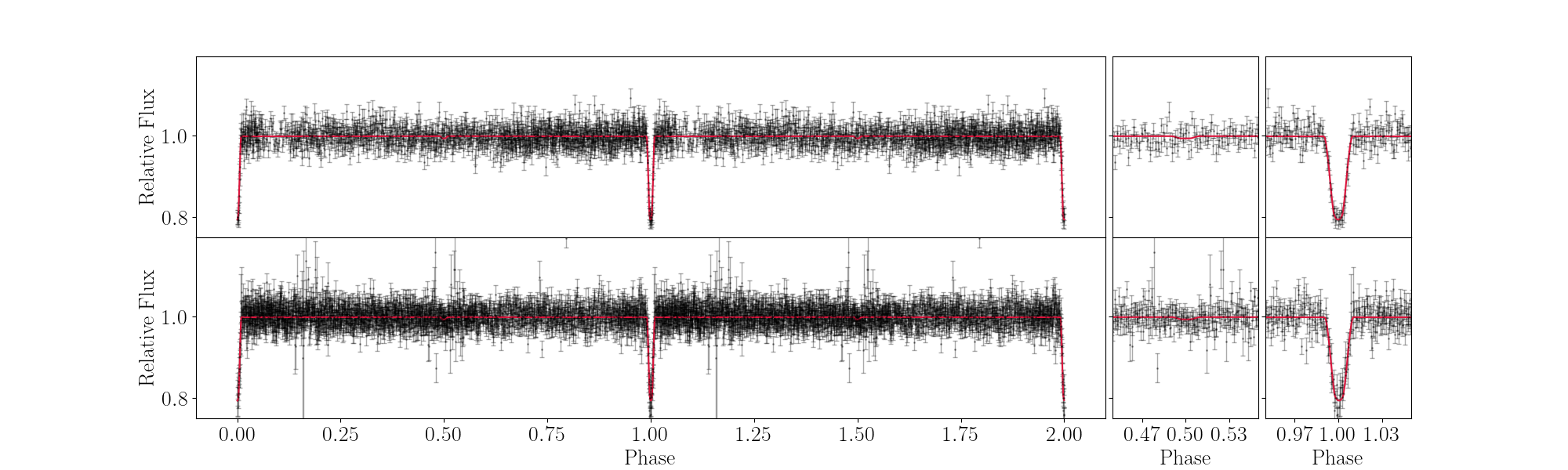}
  \caption{McDonald 2.1-meter $g$-band (top) and $r$-band (bottom) light curves of J0221$+$1710. Best-fitting \textsc{lcurve} models are overplotted in red. We provide zoomed-in subplots showing the regions surrounding the primary and secondary eclipses.}
  \label{fig:0221p1710_lcurve}
\end{figure*}

\begin{table}
\center
  \renewcommand{\arraystretch}{1.5}
  \addtolength{\tabcolsep}{2pt}
    \hspace*{-1.5cm}
	\begin{tabular}{l c c c}
    \hline
    \hline
    {} & \multicolumn{1}{C}{\rm J0221$+$1710} & \multicolumn{1}{C}{\rm J1812$+$0525} & \multicolumn{1}{C}{\rm J2049$+$3351}\\
    \hline
    {$q$} & \nodata & {$0.41\pm0.03$} & $0.39\pm0.2$ \\
    {$i~({\rm ^\circ})$} & $89.0\pm0.2$ & {$75^{+2}_{-5}$} & $74^{+3}_{-2}$ \\
    {$R_1/a$} & $0.045\pm0.001$ & {$0.20\pm0.01$} & $0.31\pm0.05$ \\
    {$R_2/a$} & $0.020\pm0.001$ & \nodata & $0.06^{+0.03}_{-0.02}$ \\
    {$T_{\rm eff,1}~({\rm K})$} & $13400\pm180$ & {$8900\pm100$} & $23200^{+4900}_{-4300}$ \\
    {$T_{\rm eff,2}~({\rm K})$} & $\lesssim6400$ & {\nodata} & $35400^{+9700}_{-8600}$ \\
    \hline
    \hline
	\end{tabular}
    \caption{System parameters for J0211$+$1710, J1812$+$0525, and J2049$+$3351, obtained through light curve modeling with \textsc{lcurve} as described in the text.}
    \label{table:lcurve_table}
\end{table}

We modeled the geometry of the binary by simultaneously fitting the $g$- and $r$-band light curves using \textsc{lcurve} \citep{copperwheat2010}, fitting for the component radii ($r_i=\frac{R_i}{a}$) and the temperature of the companion ($T_{\rm eff,2}$). We fixed the orbital period to the value obtained from our ZTF BLS analysis and used the results from our spectroscopic follow-up ($T_{\rm eff,1}$, $\log{g_1}$, and $K_1$) as Gaussian priors to our light curve modeling. We interpolated over the grid of gravity darkening and quadratic limb darkening coefficients from \citet{claret2020} for our primary star based on our spectroscopic values and used values for a companion with $T_{\rm eff,2}=10{\rm ,}000~{\rm K}$ and $\log{g_2}=8.0$.

The most probable system parameters from our light curve fitting are $R_1=0.028\pm0.001~{\rm R_\odot}$, $R_2=0.012\pm0.001~{\rm R_\odot}$, $i=89.0\pm0.2^\circ$, and $T_{\rm eff,2}=5200^{+400}_{-500}~{\rm K}$. However, given the quality of our McDonald $r$-band light curve data during the eclipse, we adopt the 3-sigma upper limit to the temperature of the companion. Our most probable model parameters are summarized in Table \ref{table:lcurve_table}. Figure \ref{fig:0221p1710_lcurve} displays our phase-folded McDonald 2.1-meter $g$- (top) and $r$-band (bottom) light curves with the best-fitting model over-plotted in red. 

With the inclination and radial velocity semi-amplitude known, we used the binary mass function to calculate the companion mass $M_2=0.58\pm0.02~{\rm M_\odot}$, corresponding to radius $R_2=0.0129~{\rm R_\odot}$, in agreement with the radius estimate from our light curve modeling.   

We similarly fit our eclipsing $g$- and $r$-band light curves for J0221$+$1710 separately using \textsc{jktebop} \citep{southworth2004,southworth2013} to confirm the consistency of our \textsc{lcurve} solution. We fit for the sum and ratio of the component radii, the inclination, and the surface brightness ratio. We performed 10,000 Monte Carlo iterations, which returned a solution degenerate in inclination ($88.6\pm0.2^\circ$ and $87.9\pm0.2^\circ$). The most probable degenerate solution agrees well with our \textsc{lcurve} results, returning: $R_1=0.029\pm0.001~{\rm R_\odot}$, $R_2=0.013\pm0.001~{\rm R_\odot}$, $i=88.6\pm0.2^\circ$, and $T_{\rm eff,2}\approx6200~{\rm K}$  We report our simultaneous $g$- and $r$-band solution from the \textsc{lcurve} solution as the true system parameters of J0221$+$1710.

With the individual component masses, orbital period, and orbital inclination known, we estimated the gravitational wave strain using the equation

\begin{equation}
    \label{eqn:strain}
    h_c=3.4\times10^{-23}\frac{\mathcal{M}^{5/3}\sqrt{\cos^4{i}+2\cos^2{i}+1}}{P^{2/3}~d},
\end{equation}

where $\mathcal{M}$ is the chirp mass, $P$ is the period in days, and $d$ is the distance in kpc \citep{timpano2006, roelofs2007}. We multiplied by $\sqrt{(4~{\rm yr})f_{\rm GW}}$ to account for the increased signal after a 4-year LISA mission. Our estimated gravitational wave strain for J0221$+$1710 is $h_c=(2.87\pm0.14)\times10^{-20}$. Additionally, we estimated the orbital decay due to gravitational wave emission using the equation

\begin{equation}
    \dot{P}=\frac{96}{5}\left(\frac{G\mathcal{M}}{P}\right)^{5/3}\frac{2\pi}{c^5},
\end{equation}

resulting in $\dot{P}=(3.86\pm0.14)\times10^{-14}~{\rm s~s^{-1}}$, which corresponds to an eclipse timing offset of $\Delta T_0\approx-3.6\pm0.1~{\rm s}$ after 10 years.
\subsection*{\rm J1121$+$6052}
Our 26 radial velocity measurements of J112157.163$+$605210.265 (J1121$+$6052; GAIA DR3 861011995046220544) return best-fitting orbital parameters $P_{\rm RV}=2.0283\pm0.0003~{\rm h}$, $K=183.5\pm2.6~{\rm km~s^{-1}}$, and $\gamma=-15.7\pm2.2~{\rm km~s^{-1}}$. We calculated the Galactic space velocities $[U,V,W]=[56.7\pm1.4, -10.2\pm1.5, 23.4\pm1.4]~{\rm km~s^{-1}}$, consistent with a short-period binary in the disk.

Our pure-hydrogen atmosphere fits to the summed zero-velocity spectrum of J1121$+$6052 resulted in atmospheric parameters $T_{\rm eff}=11{\rm ,}690\pm170~{\rm K}$ and $\log{g}=5.41\pm0.05$, which suggests white dwarf mass $M_1=0.19\pm0.01~{\rm M_{\odot}}$ based on the ${\rm Z}=0.02$ model tracks of \citet{istrate2016}.

We find a weak periodic signal in the TESS 2-minute cadence data through our Lomb-Scargle periodicity search of J1121$+$6052 (TIC 417868394). The TESS light curve shows periodic variability at the 1\% level with frequency $f_{\rm TESS}=23.6649~{\rm cycles~d^{-1}}$ ($P_{\rm TESS}=1.0142~{\rm h}$); half the period obtained through radial velocity measurements. We display the phase-folded TESS 2-minute cadence light curve in Figure \ref{fig:combined_tess} (upper-right). J1121$+$6042 is in a relatively isolated field, with no nearby bright stars to heavily dilute the TESS light curve. We do not recover this weak periodic signal in the available ground-based ZTF data.

\subsection*{\rm J1506$-$1125}

\begin{figure}
  \includegraphics[scale=0.6]{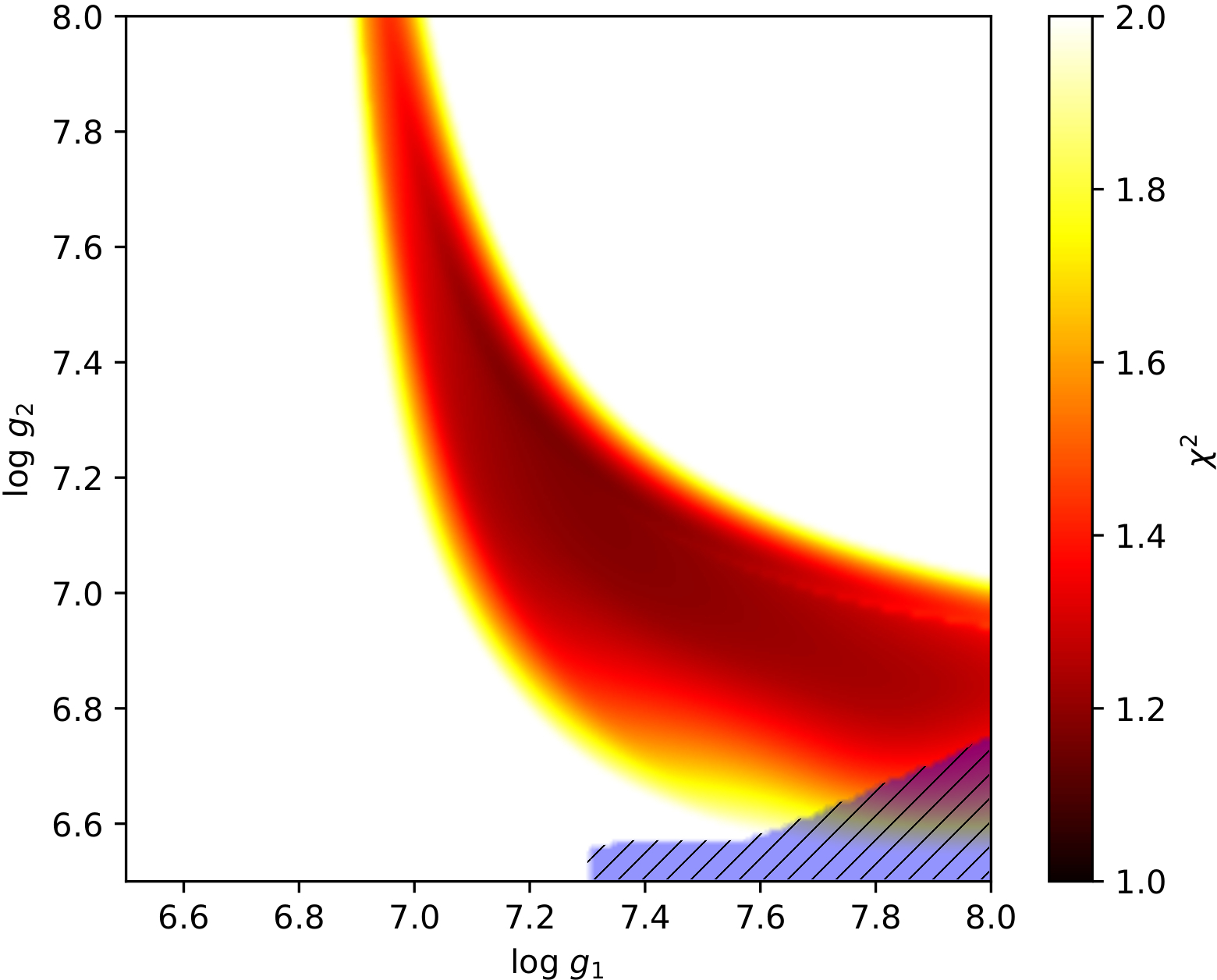}
  \caption{Normalized $\chi^2$ distribution from our fits to the combined SED and optical spectroscopy for J1506$-$1125, as a function of $\log{g_1}$ and $\log{g_2}$. The hatched blue region represents the region of parameter space excluded by constraints from the binary mass function.}
  \label{fig:1506_chi2_plot}
\end{figure}

\begin{figure*}
  \centerline{\includegraphics[scale=0.60]{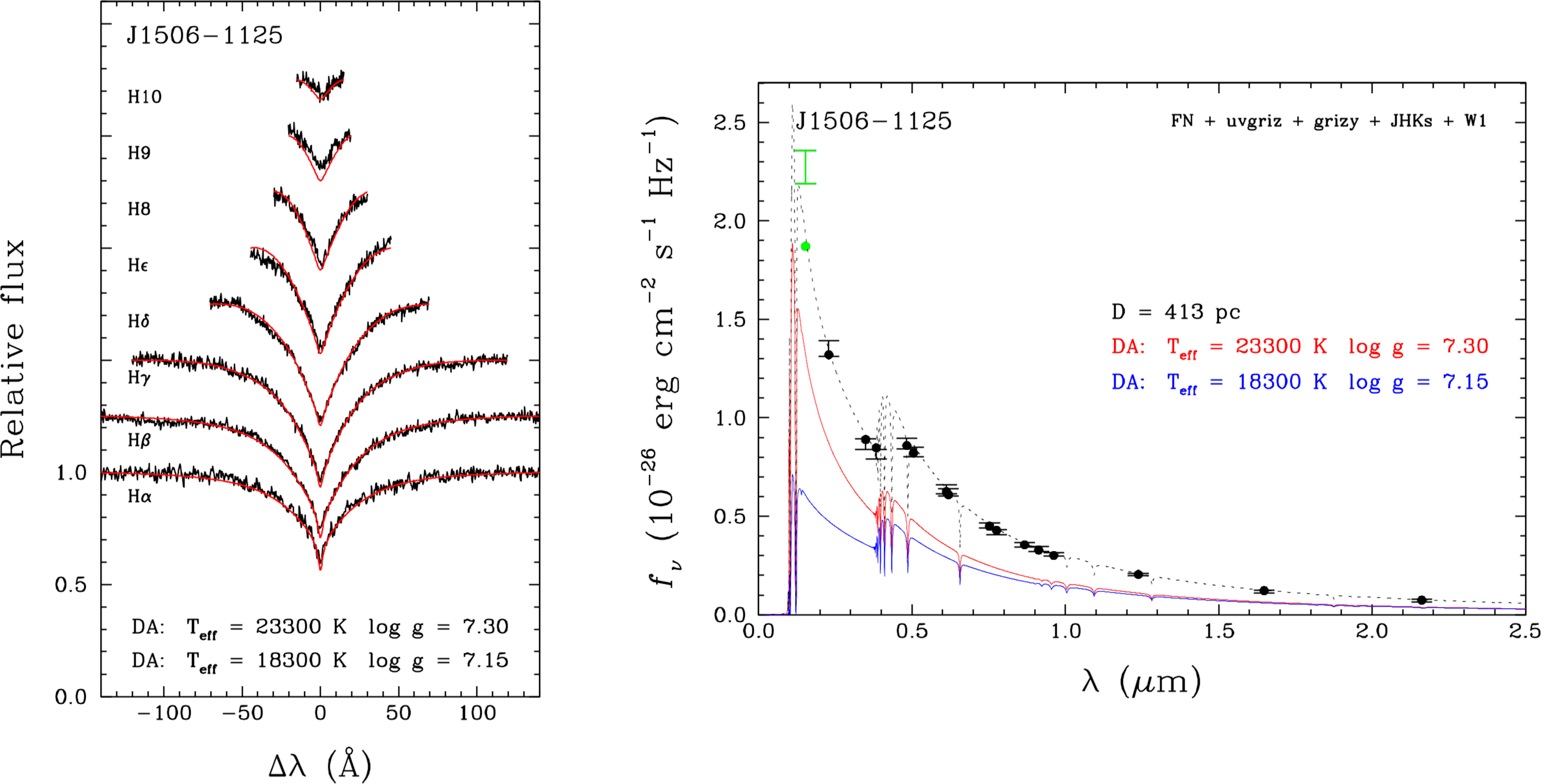}}
  \caption{Example model-atmosphere fit to the spectrum (left) and SED (right) of J1506$-$1125 including contributions from two DA white dwarfs. In the left panel, the observed and predicted Balmer lines are shown as black and red lines, respectively. In the right panel, the observed and predicted average fluxes are displayed as error bars and filled circles, respectively; for reference, the red and blue lines show the individual contributions of the components to the total monochromatic model flux, which is displayed as a black dotted line. The GALEX FUV photometry measurement (green) is excluded from our SED fit. This fit does not represent a unique solution (see text).}
  \label{fig:1506m1125_double_fit}
\end{figure*}

We obtained 31 radial velocity measurements for J150612.345$-$112511.994 (J1506$-$1125; Gaia DR3 6312837970697953920). Our best-fitting circular orbit fits suggest orbital parameters $P_{\rm RV}=0.7757\pm0.0094~{\rm h}$, $K=167.5\pm4.3~{\rm km~s^{-1}}$, and $\gamma=43.5\pm2.8~{\rm km~s^{-1}}$. J1506$-$1125 is likely a disk object with Galactic space velocities $[U,V,W]=[41.8\pm1.6,-7.3\pm1.5,13.3\pm1.4]~{\rm km~s^{-1}}$.

Our pure-hydrogen model atmosphere fits to the summed zero-velocity spectrum result in best-fitting atmospheric parameters $T_{\rm eff}=22{\rm ,}050\pm320~{\rm K}$ and $\log{g}=7.44\pm0.05$, which corresponds to a white dwarf with mass $M_1=0.43\pm0.02~{\rm M_{\odot}}$ based on disk-metallicity models of \cite{althaus2013}. Our radial velocity measurements suggest a minimum companion mass $M_{\rm 2,min}=0.18\pm0.01~{\rm M_\odot}$, significantly below the mass of the visible star.

Interestingly, our distance estimates suggest that J1506$-$1125 is over-luminous when compared to our single-star SED models, which suggests our single-star atmospheric model parameters may be inaccurate due to significant contribution to the system light from a companion. However, given the relatively low radial velocity semi-amplitude ($K=167.5\pm4.3~{\rm km~s^{-1}}$), we are not able to resolve individual absorption components in our Magellan 6.5-meter optical spectrum with $\approx1.0~{\rm \AA}$ resolution, if they are present.

We performed simultaneous model atmosphere fits to the SED and median-combined optical spectrum of J1506$-$1125, including contribution from two components in a binary \citep[see][for the details of the method]{bedard2017,kilic2020}. The SED was built from the available GALEX \cite[NUV;][]{martin2005}, SkyMapper \citep[uvgriz;][]{onken2019}, Pan-STARRS \citep[grizy;][]{chambers2016}, 2MASS \citep[JH${\rm K_s}$;][]{skrutskie2006}, and AllWISE \citep[W1;][]{wright2010,mainzer2011} photometry, and dereddened using the extinction maps of \citet{schlafly2011}. Figure \ref{fig:1506_chi2_plot} shows the resulting $\chi^2$ distribution plot as a function of $\log{g_1}$ and $\log{g_2}$, with the region of parameter space excluded by constraints from the binary mass function shaded in blue. Dark regions indicate the regions with the most probable system parameters based on our fits.

Because we do not resolve individual absorption components in our optical spectrum, we are unable to identify a unique solution to describe the atmospheric parameters of the components of J1506$-$1125. Figure \ref{fig:1506m1125_double_fit} displays an example fit near the center of the dark region, defined by $T_{\rm eff,1}=23{\rm ,}300~{\rm K}$, $\log{g_1}=7.30$, $T_{\rm eff,2}=18{\rm ,}300~{\rm K}$ and $\log{g_2}=7.15$, corresponding to masses $M_1=0.40~{\rm M_\odot}$ and $M_2=0.33~{\rm M_\odot}$ based on He-core cooling tracks of \citet{althaus2013}. However, the extremes of our probable parameter space allow for solutions at [($T_{\rm eff,1}$,$T_{\rm eff,2}$), ($\log{g_1}$,$\log{g_2}$), ($M_1$,$M_2$)] = [($22{\rm ,}000~{\rm K}$, $18{\rm ,}100~{\rm K}$), ($7.05$, $7.55$), ($0.34~{\rm M_\odot}$, $0.43~{\rm M_\odot}$)] and [($27{\rm ,}300~{\rm K}$, $17{\rm ,}900~{\rm K}$), ($7.80$, $6.90$), ($0.54~{\rm M_\odot}$, $0.28~{\rm M_\odot}$)], where we used C/O-core cooling tracks of \citet{bedard2020} for the relatively massive $0.54~{\rm M_\odot}$ solution.

We find no TESS Full-Frame image data, 2-minute cadence data, or 20-second cadence data for J1506$-$1125 up to sector 56. Our Lomb-Scargle and BLS algorithms do not identify periodic photometric variability in the public ZTF DR13 data archive.

\subsection*{\rm J1812$+$0525}
The orbital period of J181238.471$+$052529.868 (J1812$+$0525; Gaia DR3 4471573464995153280) was constrained through its periodic photometric variability identified in our search of the public ZTF data archive. Our Lomb Scargle periodogram identified periodic photometric variability at $P_{\rm ZTF}=43.1~{\rm min}$.

\begin{figure*}
  \center
  \includegraphics[scale=0.2]{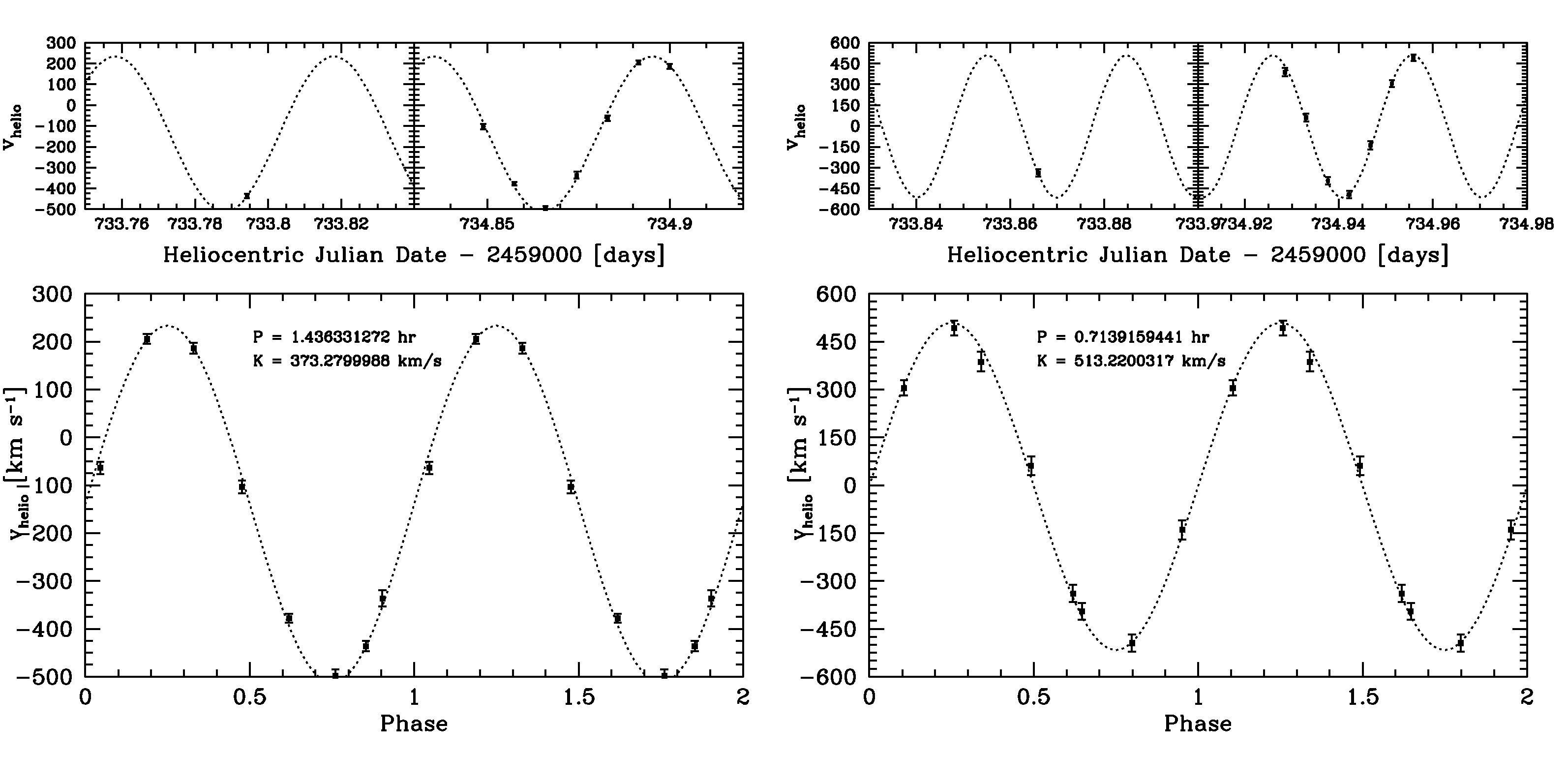}
  \caption{Orbital solutions for J1812$+$0525 (left) and J2049$+$3351 (right).}
  \label{fig:combined_rv}
\end{figure*}

We obtained a single follow-up spectrum on 2022 June 03 to confirm its nature. We then completed its orbital solution with seven additional back-to-back spectra during the following night. Figure \ref{fig:combined_rv} (left) displays our orbital solution for J1812$+$0525. Our best-fitting radial velocity solution finds velocity semi-amplitude $K=373.3\pm6.2~{\rm km~s^{-1}}$, systemic velocity $\gamma=-139.6\pm4.7~{\rm km~s^{-1}}$, and orbital period $P_{\rm RV}=1.436\pm0.002~{\rm h}$ ($86.2\pm0.1~{\rm min}$), in good agreement with the half-period identified from the ZTF data. We use the precise Gaia DR3 astrometry, together with our radial velocity information, to estimate its Galactic velocities $[U,V,W]=[-92.9\pm2.3,-87.4\pm2.8, -1.1\pm2.4]~{\rm km~s^{-1}}$, which suggests that J1812$+$0525 is a short-period binary in the Galactic Halo.

We fit the summed zero-velocity spectrum with pure-hydrogen model atmospheres and obtained best-fitting atmospheric parameters $T_{\rm eff,0}=9300\pm120~{\rm K}$ and $\log{g_0}=6.11\pm0.05$. Because this is a cool object, we corrected its atmospheric solution for 3D effects using equations provided in \citet{tremblay2015}, resulting in corrected atmospheric parameters $T_{\rm eff}=8960\pm130~{\rm K}$ and $\log{g}=5.96\pm0.08$, corresponding to a primary mass $M_1=0.28\pm0.03~{\rm M_{\odot}}$ based on the halo-metallicity models tracks of \citet{istrate2016}.

We used the McDonald Observatory 2.1-meter telescope to confirm photometric variability at $P=43.4~{\rm min}$, roughly consistent with the half-period seen in the public ZTF data. However, our data quality is too poor to reasonably constrain orbital parameters through photometric fitting.

We instead modelled the ZTF DR16 $g$- and $r$-band data simultaneously using \textsc{lcurve}. Our free parameters included the mass ratio ($q=\frac{M_1}{M_2}<1.0$), scaled primary star radius ($r_1=\frac{R_1}{a}$), orbital inclination ($i$), primary star effective temperature ($T_{\rm eff,1}$), time of superior conjunction ($t_0$), and the velocity scale (the sum of the unprojected orbital speeds). We fixed the gravity and quadratic limb-darkening coefficients to values from \citet{claret2020} for a $T_{\rm eff,1}=9000~{\rm K}$, $\log{g_1}=6.0$ primary with a $T_{\rm eff,2}=5500~{\rm K}$, $\log{g_1}=8.0$ companion and assign Gaussian priors based on our spectroscopic values for $T_{\rm eff,1}$, $\log{g_1}$, and $K_1$. The most-probable system parameters based on our light curve modelling are $q=0.42\pm0.03$, $r_1=0.20\pm0.01$, $i=75^{+2}_{-5}~^\circ$, and $T_{\rm eff,1}=8900\pm100~{\rm K}$. These parameters are summarized in Table \ref{table:lcurve_table}. With constraints from our radial velocity measurements, this corresponds to an $M_1=0.30\pm0.03~{\rm M_\odot}$ primary star with an $M_2=0.73^{+0.05}_{-0.04}~{\rm M_\odot}$ companion.

\subsection*{\rm J2049$+$3351}
The orbital period of J204951.274$+$335153.126 (J2049$+$3351; Gaia DR3 1869111286948848128) was constrained through its periodic photometric variability identified in our search of the public ZTF data archive. Our Lomb Scargle algorithm identified strong variability at $P_{\rm ZTF}=21.6~{\rm min}$ with amplitude $A\approx0.07~{\rm mag}$.

We obtained a single follow-up spectrum on 2022 June 03 to confirm its nature and completed its orbital solution with seven additional back-to-back spectra on the following night. Figure \ref{fig:combined_rv} (right) displays our orbital solution for J2049$+$3351. Our best-fitting radial velocity solution yields velocity semi-amplitude $K=513.2\pm9.5~{\rm km~s^{-1}}$, systemic velocity $\gamma=-3.4\pm7.7~{\rm km~s^{-1}}$, and orbital period $P_{\rm RV}=0.7139\pm0.0002~{\rm h}$ ($42.834\pm0.012~{\rm min}$), in good agreement with the half-period identified from the ZTF data. The Galactic space velocities, $[U,V,W]=[-9.8\pm2.4,-8.1\pm2.6, 13.4\pm2.3]~{\rm km~s^{-1}}$, place J2049$+$3351 in the Galactic disk.

\begin{figure}
  \center
  \includegraphics[scale=0.30]{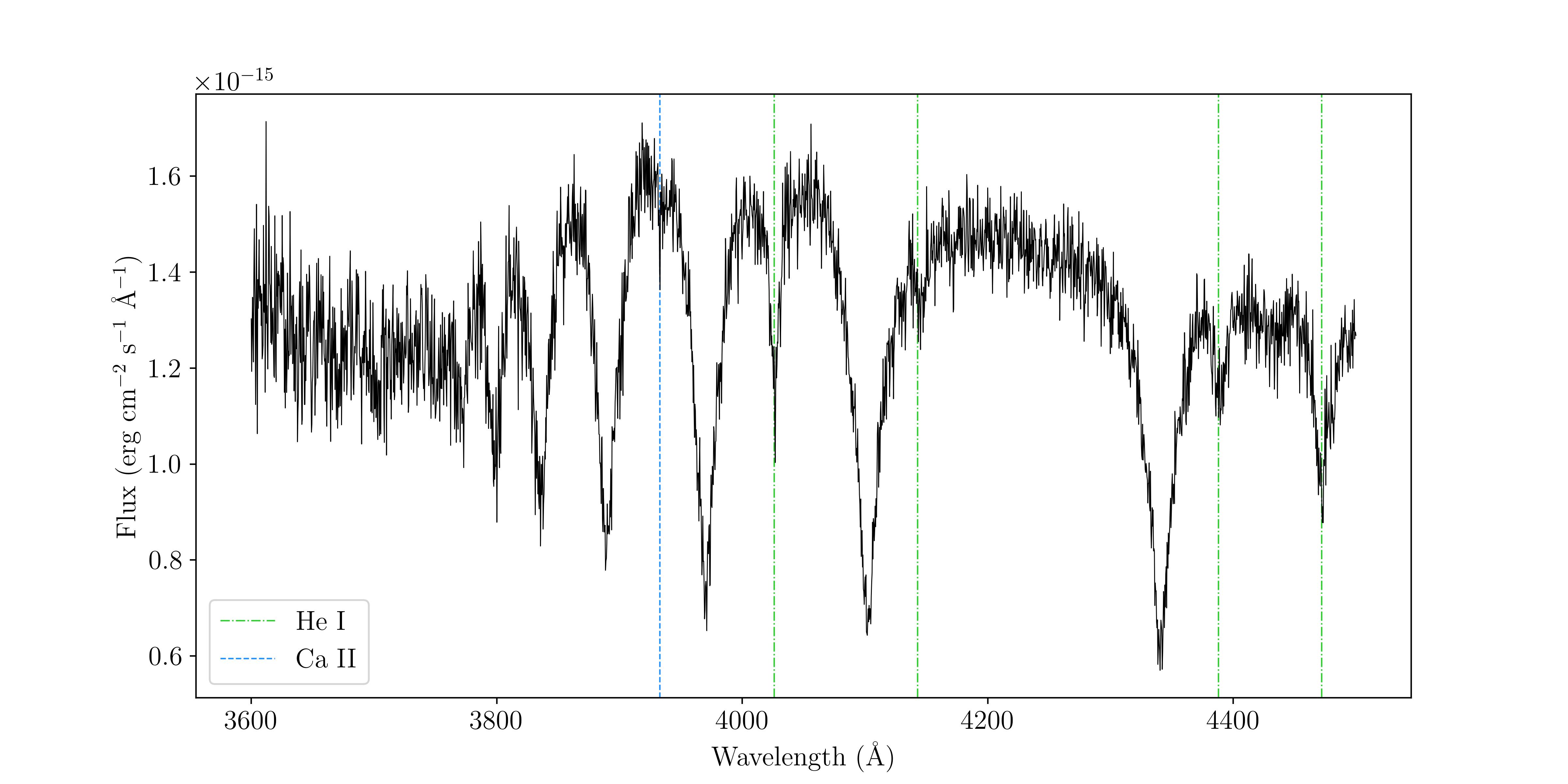}
  \caption{Co-added MMT optical spectrum of J2049$+$3351. In addition to the dominant Hydrogen Balmer absorption, strong neutral Helium absorption features (green dash-dot lines) can be seen at $4472~{\rm \AA}$, $4388~{\rm \AA}$, $4143~{\rm \AA}$, and $4026~{\rm \AA}$, as well as a weak Ca II absorption feature (blue dash lines) at $3933~{\rm \AA}$.}
  \label{fig:2049p3351_spec}
\end{figure}

J2049$+$3351 has an interesting optical absorption spectrum, including broad Hydrogen Balmer lines with strong He I absorption at $4472~{\rm \AA}$, $4388~{\rm \AA}$, $4143~{\rm \AA}$, and $4026~{\rm \AA}$, as well as a weak Ca II feature at $3933~{\rm \AA}$. Our MMT 6.5-meter optical spectrum of J2049$+$3351 is presented in Figure \ref{fig:2049p3351_spec}. We note that the individual Hydrogen and Helium absorption lines appear to move in-sync with each other throughout the orbit of the binary.

\begin{figure*}
  \includegraphics[scale=0.6]{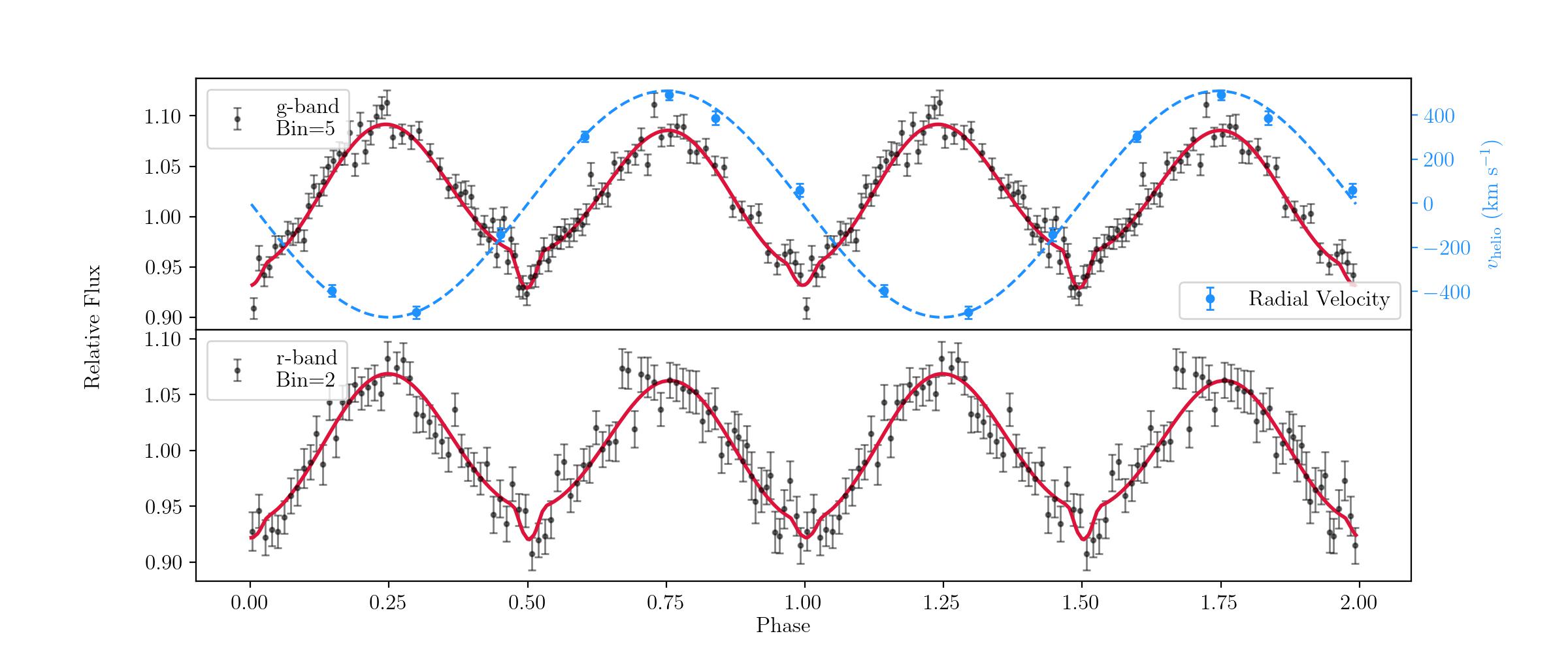}
  \caption{Top: McDonald 2.1-meter $g$-band (top; binned by 5) and $r$-band (bottom; binned by 2) light curves of ZTF J2049$+$3351. We overplot our radial velocity measurements as blue data points in the top panel.}
  \label{fig:2049p3351_lcurve_models}
\end{figure*}

\begin{figure}
  \includegraphics[scale=0.6]{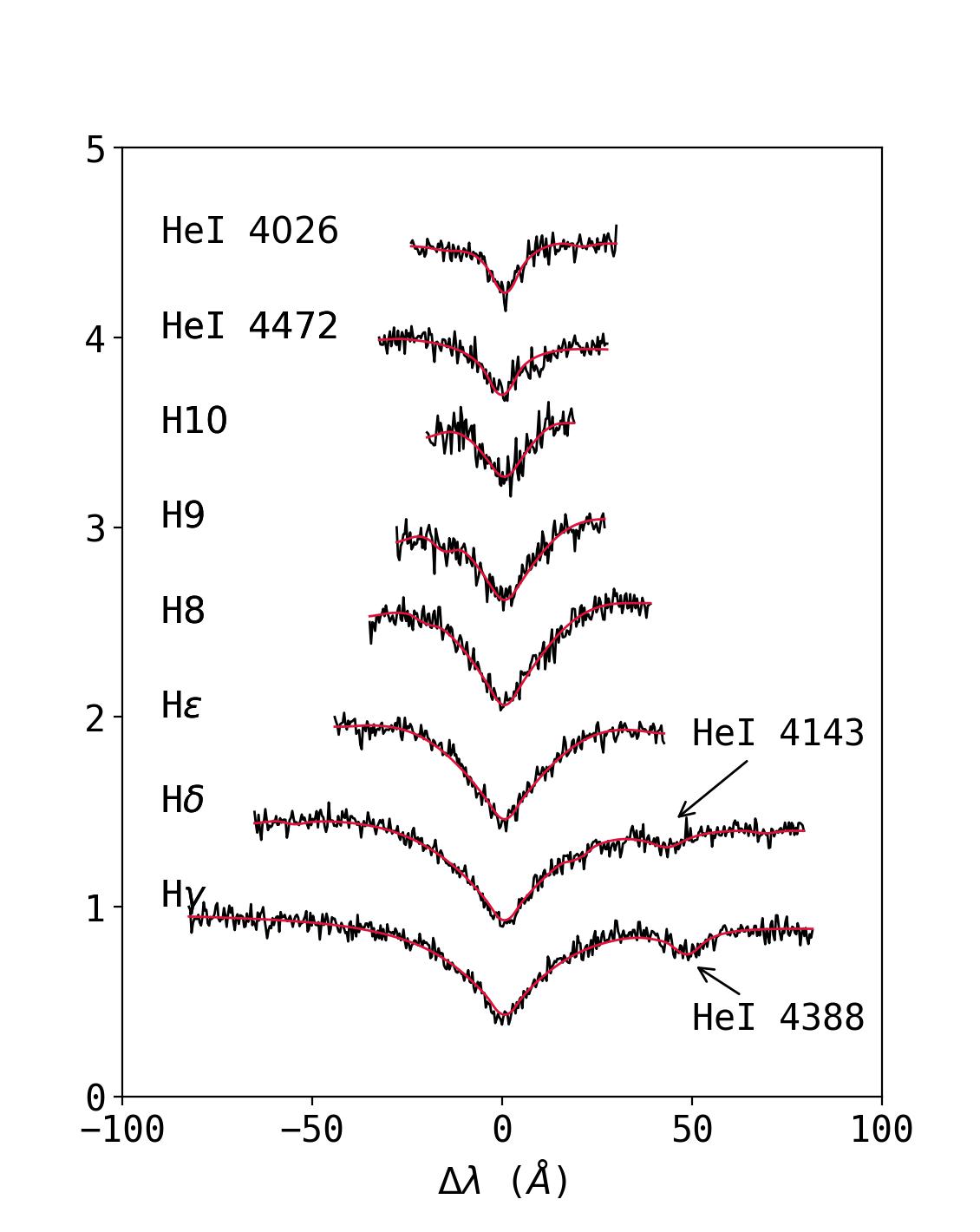}
  \caption{Best fitting spectroscopic model to the optical spectroscopy of J2049$+$3351, including rotational broadening.}
  \label{fig:2049p3351_spec_fit}
\end{figure}

To fit the co-added spectrum of J2049$+$3351, we use spectral models which were constructed using a hybrid LTE/NLTE approach described in detail in \citet{przybilla2011} and \citet{irrgang2021}. The grid of spectral models covers a typical range of hot subdwarf $T_{\rm eff}$, and $\log{g}$, up to modest helium abundances \citep{heber2023,irrgang2021, irrgang2022}. Our best-fitting parameters, with their bootstrapped uncertainties, are $T_{\rm eff}=17{\rm ,}500\pm500~{\rm K}$, $\log{g}=5.85\pm0.05$, $\log{\frac{He}{H}}=-0.24\pm0.07$, and $v_{\rm rot}=260\pm60~{\rm km~s^{-1}}$. Best-fitting solutions without rotational broadening fail to reproduce the observed broad Helium absorption lines. We present the best-fitting model atmosphere over-plotted on our optical spectrum in Figure \ref{fig:2049p3351_spec_fit}.

We obtained $218~{\rm min}$ of $g$-band and $108~{\rm min}$ of $r$-band high-speed photometry using the McDonald 2.1-meter telescope. We performed a simultaneous multi-band fit to our unbinned light curves using \textsc{lcurve}. We assign the time of primary conjunction using our radial velocity data and fit for the mass ratio, individual stellar radii, orbital inclination, and effective temperatures. 

Our best-fitting model identifies primary and secondary eclipses in the data. However, J2049$+$3351 is faint (${\rm Gaia}~G=18.7~{\rm mag}$), and our 2.1-meter McDonald light curves are noisy. While the eclipses are not seen in our $r$-band light curve due to the high noise level, a relatively clean secondary-eclipse feature can be seen in the binned $g$-band light curve data and a noisy primary-eclipse feature may be visible in our unbinned $g$-band data. We present our binned and phase-folded McDonald 2.1-meter $g$- and $r$-band light curves in Figure \ref{fig:2049p3351_lcurve_models} with the most-probable light curve model over-plotted as a red line and radial velocity curve over-plotted in blue.

Our best-fitting model to the McDonald 2.1-meter light curves finds mass ratio $q=\frac{M_1}{M_2}=0.39\pm0.2$, inclination $i=74\pm3^\circ$, $T_{\rm eff,1}=23{\rm ,}200^{+4900}_{-4300}~{\rm K}$,$T_{\rm eff,2}=35{\rm ,}400^{+9700}_{-8600}~{\rm K}$, $r_1=\frac{R_1}{a}=0.31\pm0.05$, and $r_2=\frac{R_2}{a}=0.06^{+0.03}_{-0.02}$. We summarize these values in Table \ref{table:lcurve_table}.

Because J2049$+$3351 is faint, large aperture high-speed photometry will be required to obtain sufficient signal-to-noise to confirm the presence of eclipses. These future observations will be used to place precise constraints on the binary parameters and determine the evolutionary history of J2049$+$3351.

\subsection*{\rm J2102$-$4145}
\begin{figure}
  \includegraphics[scale=0.40]{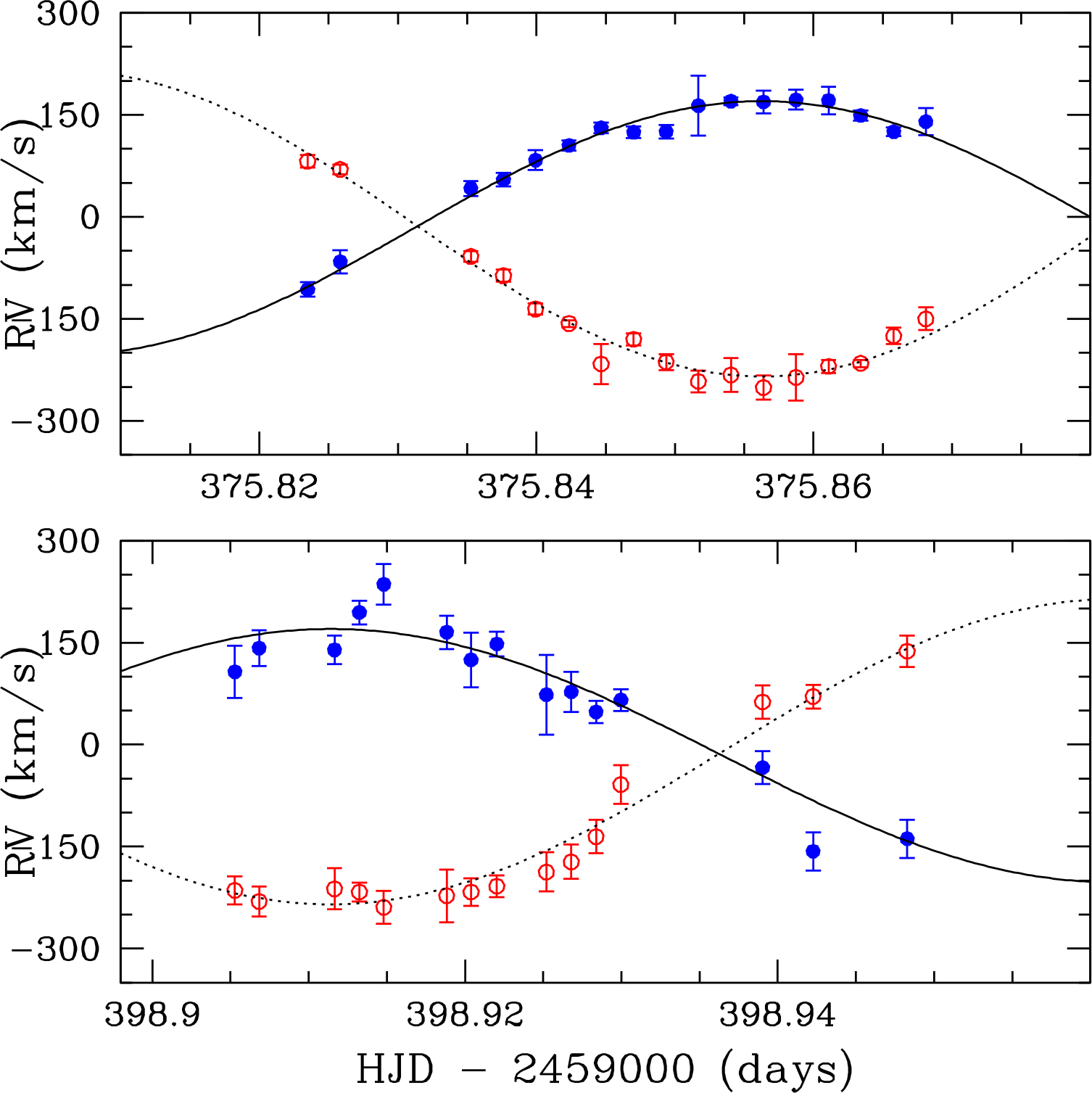}
  \caption{Orbital solutions for each component of the double-lined binary J2102$-$4145. Top: Magellan 6.5-meter telescope data. Bottom: Gemini South 8.1-meter telescope data. The blue/red points represent the radial velocity measurements for the primary/secondary star.}
  \label{fig:2102m4145_double_rv}
\end{figure}

J210220.456$-$414501.736 (J2102$-$4145; Gaia DR3 6581249825853801984) is a relatively bright (Gaia ${\rm G}=15.7~{\rm mag}$) and nearby ($d_\pi=165\pm1~{\rm pc}$) eclipsing low mass white dwarf binary. Our 6.5-meter Magellan spectroscopic follow-up observations reveal two nearly-equal depth absorption components, most easily seen in ${\rm H}\alpha$. Thus J2102$-$4145 is now the eighth confirmed double-lined, double-degenerate, eclipsing white dwarf binary published after CSS41177 \citep[J1005$+$2249;][]{bours2014}, ZTF J1539$+$5027 \citep{burdge2019}, ZTF J1901+5309 \cite{coughlin2020}, and ZTF J0538$+$1953, ZTF J0722$-$1839, ZTF J1749$+$0924, and ZTF J2029$+$1534 \citep{burdge2020}.

Because the individual absorption components are roughly equal-depth, obtaining a precise orbital period and individual velocity semi-amplitudes leads to significant period aliases when combining data over multiple nights. Our 55 radial velocity measurements for J2102$-$4145 provide orbital period constraints with strong period aliases between $P_{\rm RV}=2-3~{\rm h}$.

Fortunately, the TESS 2-minute cadence data archive provides a light curve for J2102$-$4145 (TIC 115013365). Our BLS algorithm identifies periodic eclipses in the TESS light curve with period $P_{\rm BLS}=1.2~{\rm h}$, with its phase-folded light curve showing a shallow primary eclipse heavily diluted by a bright (Gaia ${\rm G}=11.2~{\rm mag}$) field star roughly $30\arcsec$ away. Figure \ref{fig:combined_tess} (lower-left) shows the phase-folded TESS 2-minute cadence light curve and its BLS periodogram for J2102$-$4145.

Given that this is a double-lined spectroscopic binary with nearly-equal depth absorption components, our BLS algorithm will confuse the nearly-equal depth primary and secondary eclipses at the true orbital period with one primary eclipse occurring twice as often. Our BLS algorithm has identified the half-period; the true orbital period of J2102$-$4145 is therefore $P=2.4~{\rm h}$, in agreement with the constraints from our radial velocity analysis.

Using the TESS period, we were able to constrain the orbital solutions for both stars in the binary. Figure \ref{fig:2102m4145_double_rv} displays our best-fitting orbital solution for each star, using data from Magellan and Gemini South. We find $P_{\rm RV}=144.3^{+0.4}_{-0.3}~{\rm min}$, $K_1=227\pm8~{\rm km~s^{-1}}$, $\gamma_1=-8^{+7}_{-6}~{\rm km~s^{-1}}$, $K_2=186\pm8~{\rm km~s^{-1}}$, and $\gamma_2=-19^{+6}_{-7}~{\rm km~s^{-1}}$, corresponding to binary mass ratio $q=0.82\pm0.05$. 

\begin{figure*}
  \centerline{\includegraphics[scale=0.60]{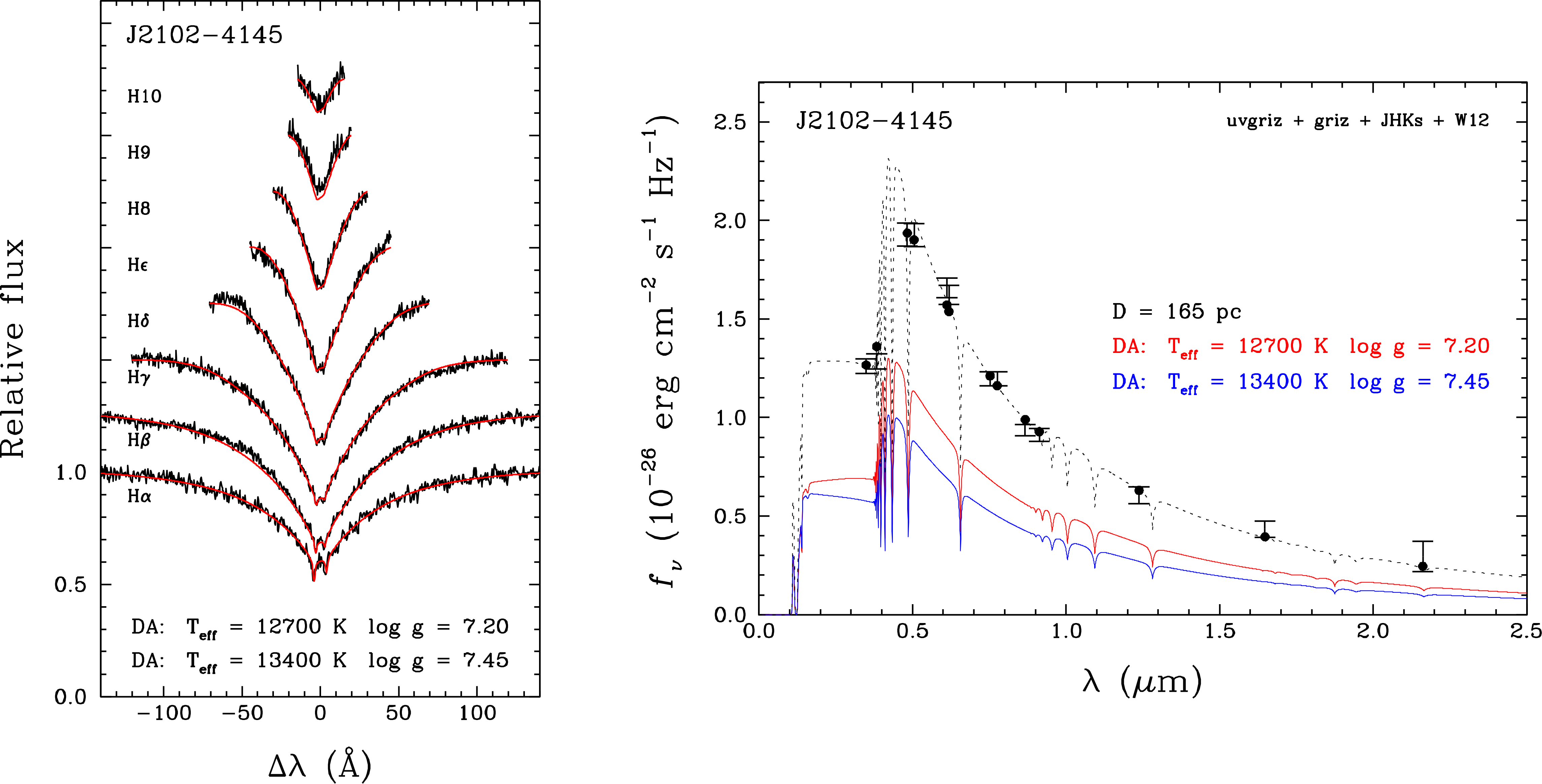}}
  \caption{Best simultaneous model-atmosphere fit to the spectrum (left) and SED (right) of J2102$-$4145 including contributions from two DA white dwarfs. In the left panel, the observed and predicted Balmer lines are shown as black and red lines, respectively. In the right panel, the observed and predicted average fluxes are displayed as error bars and filled circles, respectively; for reference, the red and blue lines show the individual contributions of the components to the total monochromatic model flux, which is displayed as a black dotted line}
  \label{fig:2102m4145_double_fit}
\end{figure*}

Finally, we performed simultaneous model atmosphere fits to the SED and median-combined spectrum of J2102$-$4145, including the contributions of both components. To avoid significant smearing, our median combined spectrum used four consecutive spectra covering 9\% of the orbit, approaching maximum separation. Our fits also made use of the Gaia DR3 parallax measurement and the mass ratio constraint provided by the orbital solution, thanks to which we were able to obtain a unique solution, shown in Figure \ref{fig:2102m4145_double_fit}.

Our best-fitting solution suggests that J2102$-$4145 contains two similar DA white dwarfs with atmospheric parameters $T_{\rm eff,1}=12{\rm ,}700\pm500~{\rm K}$, $\log{g}_1=7.20\pm0.05$, $T_{\rm eff,2}=13{\rm ,}400\pm500~{\rm K}$, and $\log{g}_2=7.45\pm0.05$, corresponding to masses $M_1=0.32~{\rm M_\odot}$ and $M_2=0.39~{\rm M_\odot}$ based on the He-core white dwarf tracks of \citet{althaus2013}, in agreement to within $2\sigma$ of the expected values based on the difference in systemic velocities measured from our radial velocity solution.

High-speed light curve follow-up of J2102$-$4145 will allow for a rare opportunity to directly measure the radii of the individual stars in the binary, independent of model estimates.

\subsection*{\rm J2303$-$2615}
We obtained 34 radial velocity measurements of J230323.542$-$261459.917 (J2303$-$2614; Gaia DR3 2382531303846872448), resulting in a best-fitting circular orbit with parameters $P_{\rm RV}=2.8367\pm0.0008~{\rm h}$, $K=302.9\pm2.3~{\rm km~s^{-1}}$ and $\gamma=-17.1\pm2.1~{\rm km~s^{-1}}$. J2303$-$2614 is a disk object with $[U,V,W]=[31.1\pm1.2,-60.0\pm1.4,18.8\pm1.4]~{\rm km~s^{-1}}$.

We fit the median-combined zero-velocity spectrum with pure-hydrogen atmosphere models and obtained best-fitting atmospheric parameters $T_{\rm eff}=11{\rm ,}280\pm170~{\rm K}$ and $\log{g}=5.43\pm0.05$, corresponding to a white dwarf with mass $M_1=0.18\pm0.01~{\rm M_{\odot}}$.
 
Our Lomb-Scargle periodogram identified strong periodic photometric variability in the TESS 2-minute cadence data of J2303$-$2614 (TIC 12970429) at frequency $f_{\rm TESS}=16.9392~{\rm cycles~d^{-1}}$ ($P=1.417~{\rm h}$), in good agreement with half the orbital period seen in our radial velocity follow-up, suggesting that J2303$-$2614 is another tidally distorted white dwarf in a compact binary. Figure \ref{fig:combined_tess} (lower-right) shows the phase-folded TESS 2-minute cadence light curve and its Lomb Scargle power spectrum. We provide a zoomed-inset plot showing a small, but significant, peak at the true orbital period of this system.

\section{Discussion}

\subsection{Detectable LISA Binaries}
General relativity predicts that the orbits of compact binaries decay due to the emission of gravitational waves. Binaries with orbital periods $P\lesssim6~{\rm h}$ will merge within a Hubble time. The shortest period white dwarf binaries, with periods $P\lesssim1~{\rm h}$, will be the dominant source of gravitational wave signal for the Laser Interferometer Space Antenna \citep[LISA;][]{amaro2017} mission, creating an incoherent noise floor in the LISA sensitivity range at mHz frequencies \citep{nelemans2001,korol2017,li2020,amaro2022}.

Over 40 compact binaries have been characterized through their electromagnetic radiation and will be individually resolved by LISA, many of which will act as verification binaries for the LISA mission data calibration \citep[see][and references therein]{finch2022,kupfer2023}. These individually resolvable gravitational wave binaries will provide a multi-messenger approach to studying compact binary evolution through their electromagnetic and gravitational wave emission. We find that two of our 28 new binaries will be detected by LISA within a 4-year mission.

Our single star spectroscopic fits to J1506$-$1125 suggests that it contains an $M_1=0.43\pm0.02~{\rm M_\odot}$ white dwarf in a $P=0.7757\pm0.0094~{\rm h}$ binary with an $M_2>0.18\pm0.01~{\rm M_\odot}$ companion at a distance of $d_\pi=413\pm18~{\rm pc}$. We used the NASA LISA Detectability Calculator\footnote{https://heasarc.gsfc.nasa.gov/lisa/lisatool/} to estimate its expected LISA signal-to-noise ratio over 4-years of observation to be ${\rm SNR}=1.9$ assuming $i=90^\circ$, or ${\rm SNR}=3.7$ assuming $i=60^\circ$. However, our single-component fits to the SED of J1506$-$1125 suggests significant contribution to the total system light from an unseen companion. Our simultaneous multi-component fits to the available SED and optical spectroscopy of J1506$-$1125 (discussed in Section \ref{sec:results}) suggest a large range of probable stellar parameters for each component. Over this range, we find signal to noise ratios between $3.9-4.7$ assuming $i=90^\circ$, or $6.1-7.3$ assuming $i=60^\circ$.

J152601.115$-$271156.660 (J1526$-$2711; Gaia DR3 6213619999912198144) is a $P=0.67\pm0.01~{\rm h}$ binary containing an $M_1=0.37\pm0.02~{\rm M_\odot}$ primary with an $M_2>0.40\pm0.02~{\rm M_\odot}$ companion at a distance $d_\pi=623\pm68~{\rm pc}$. J1526$-$2711 is a detectable LISA binary with 4-year signal-to-noise ${\rm SNR}=3.3$ assuming $i=90^\circ$, or ${\rm SNR}=6.4$ assuming $i=60^\circ$.

We find no photometric variability for either of these binaries in the public ZTF and TESS data archives. LISA detections will provide precise constraints to their orbital inclinations and chip masses, which will be used to directly constrain individual component masses, leading to estimates on the eventual merger outcomes.

\begin{figure}
  \includegraphics*[scale=0.25]{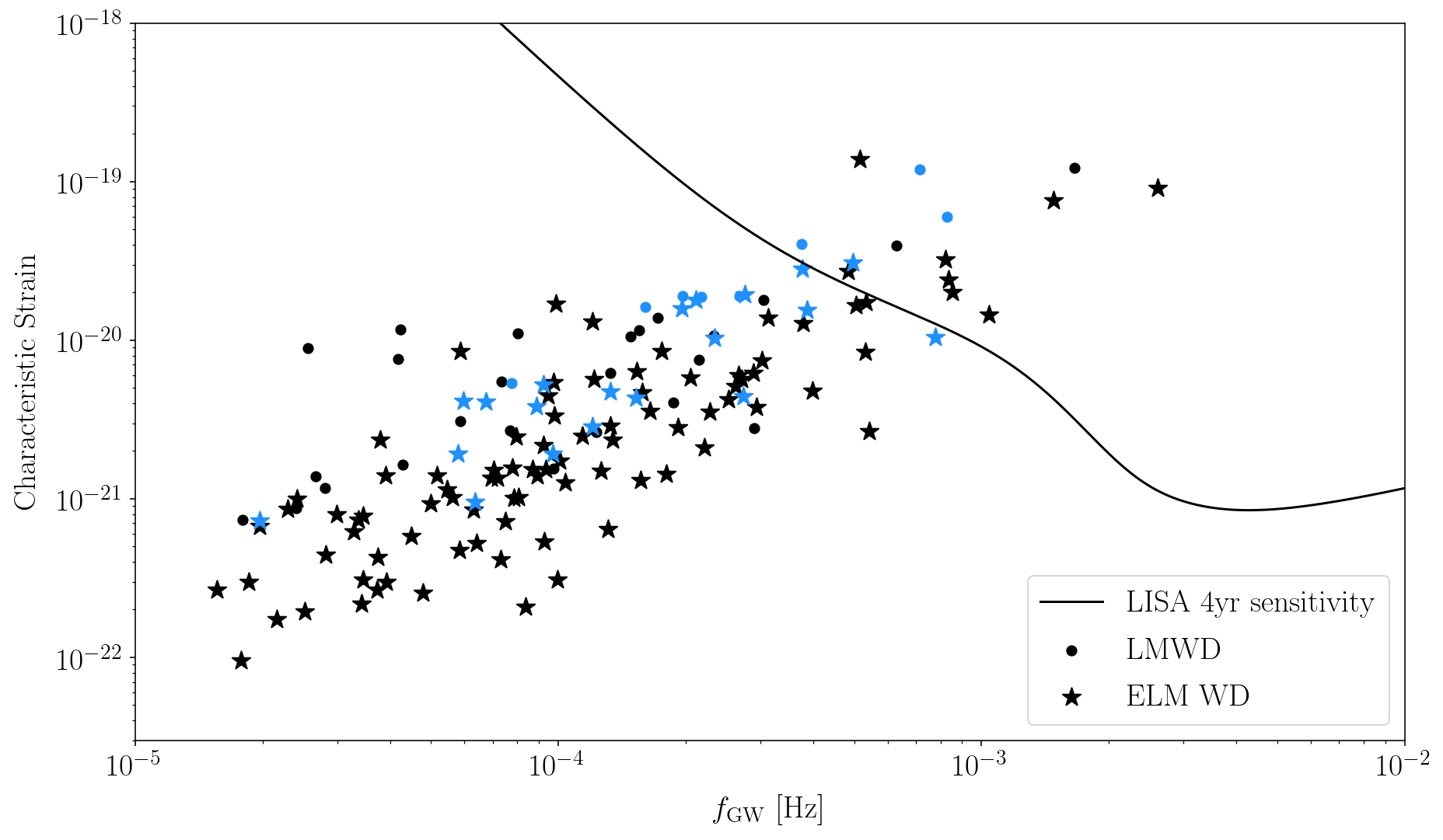}
  \caption{Four-year LISA sensitivity plot including the white dwarf binaries identified as part of the ELM Survey (black) and the 28 new white dwarf binaries from this work (blue). ELM white dwarfs ($M\leq0.305~{\rm M_\odot}$) are marked as star symbols while low-mass white dwarfs are marked as circles.}
  \label{fig:lisa_plot}
\end{figure}

Figure \ref{fig:lisa_plot} displays the LISA 4-year sensitivity curve for the ELM Survey binaries (black) with the 28 new binaries presented here (blue). We calculated the gravitational wave strain for each of our new binaries using Equation \ref{eqn:strain}, assuming $i=90^\circ$ if orbital inclination was not known through photometric constraints.

\subsection{Comparison with other surveys}
\citet{pelisoli2019} created a catalog of 5672 ELM white dwarf candidates based on Gaia DR2 astrometry with extensive target cuts to remove contamination. We cross-matched our 28 new binaries with the Gaia DR2 ELM white dwarf candidate catalog of \citet{pelisoli2019} and find thirteen matches, which we mark in Tables \ref{table:spectroscopy_table} and \ref{table:orbit_table}. Many of the remaining binaries not reported as ELM white dwarf candidates in \citet{pelisoli2019} were excluded in their color cuts used to remove cataclysmic variables and WD+M binary contaminants.

\citet{wang2022} used the LAMOST DR8 low resolution spectroscopy to further refine the candidate list of \citet{pelisoli2019} and identified 21 high probability ELM white dwarfs based on spectroscopic fits to the low resolution LAMOST data, including two of the new binaries we present in this work. \citet{wang2022} identified J0215$+$0155 with atmospheric parameters $T_{\rm eff}=10{\rm ,}540\pm40~{\rm K}$ and $\log{g}=5.06\pm0.07$, significantly lower than the values we find for our combined spectrum. Additionally, the authors identify significant radial velocity variability in two spectra of J1129$+$4715, with $T_{\rm eff,1}=11{\rm ,}670\pm50~{\rm K}$, $\log{g_1}=5.31\pm0.06$, and $v_{\rm rad,1}=-38.9\pm3.0~{\rm km~s^{-1}}$ and $T_{\rm eff,2}=11{\rm ,}150\pm90~{\rm K}$, $\log{g_2}=5.06\pm0.02$, and $v_{\rm rad,2}=94.9\pm5.4~{\rm km~s^{-1}}$, the first of which is in excellent agreement with the atmospheric parameter estimates from the fits presented in this work. 

\subsection{Conclusions}
This work has identified 28 new low mass white dwarf binaries, bringing the total number of ELM Survey binaries to 148, with 41 located in the southern sky. Interestingly, this work identifies only three halo binaries among our 28 new binaries ($\approx11\%$). Previous ELM Survey results suggest a significant contribution from halo objects of $\approx30-40\%$ \citep{gianninas2015, brown2022}, likely due to the early ELM Survey target selection being based on photometry from SDSS, which observed at high Galactic latitudes. Additionally, our low fraction of new Halo objects is likely affected by our Gaia-based selection, and in-progress object follow-up, favoring nearby objects with reliable parallax measurements, rather than more distant halo objects.

Large-scale time-domain surveys are an excellent tool for the discovery of photometrically variable systems. \citet{burdge2020} identified 15 ultra-compact ($P<1~{\rm h}$) binaries which show photometric variability in the ZTF data archive. We used the ZTF and TESS data archives to identify photometric variability in seven of our new binaries, including three short period eclipsing binaries. These short-period eclipsing binaries are especially important for determining precise physical parameters of both stars in the binary, which are valuable for binary population studies. Similar upcoming large-scale time-domain surveys, such as BlackGEM \citep{bloemen2015} and the Vera C. Rubin
Observatory's Legacy Survey of Space and Time (LSST) program \citep{ivezic2019} will enable efficient identification and characterization of white dwarf binaries in the southern sky, which will quickly expand the known population of ELM white dwarf binaries and allow for a more detailed population study of all-sky ELM white dwarf binaries.

\newpage
\section*{Acknowledgements}
AK acknowledges support from NASA through grant 80NSSC22K0338.

MK acknowledges support from NSF under grants  AST-1906379 and AST-2205736, and the NASA under grant 80NSSC22K0479.

TK acknowledges support from the National Science Foundation through grant AST \#2107982, from NASA through grant 80NSSC22K0338 and from STScI through grant HST-GO-16659.002-A. 

M.A.A.~acknowledges support from a Fulbright U.S.~Scholar grant co-funded
by the Nouvelle-Aquitaine Regional Council and the Franco-American
Fulbright Commission. M.A.A.~also acknowledges support from a Chr\'etien
International Research Grant from the American Astronomical Society.

We would like to thank Ulrich Heber and Andreas Irrgang for the providing the spectral models for J2049$+$3351.

We thank the anonymous referee for helpful comments and suggestions that improved the quality of this work.

This work has made use of data from the European Space Agency (ESA) mission
{\it Gaia} (\url{https://www.cosmos.esa.int/gaia}), processed by the {\it Gaia}
Data Processing and Analysis Consortium (DPAC,
\url{https://www.cosmos.esa.int/web/gaia/dpac/consortium}). Funding for the DPAC
has been provided by national institutions, in particular the institutions
participating in the {\it Gaia} Multilateral Agreement.

Based on observations obtained at the international Gemini Observatory, a program of NSF's NOIRLab, which is managed by the Association of Universities for Research in Astronomy (AURA) under a cooperative agreement with the National Science Foundation on behalf of the Gemini Observatory partnership: the National Science Foundation (United States), National Research Council (Canada), Agencia Nacional de Investigaci\'{o}n y Desarrollo (Chile), Ministerio de Ciencia, Tecnolog\'{i}a e Innovaci\'{o}n (Argentina), Minist\'{e}rio da Ci\^{e}ncia, Tecnologia, Inova\c{c}\~{o}es e Comunica\c{c}\~{o}es (Brazil), and Korea Astronomy and Space Science Institute (Republic of Korea).

Based on observations obtained at the Southern Astrophysical Research (SOAR) telescope, which is a joint project of the Ministério da Ciência, Tecnologia e Inovações do Brasil (MCTI/LNA), the US National Science Foundation’s NOIRLab, the University of North Carolina at Chapel Hill (UNC), and Michigan State University (MSU).

Observations reported here were obtained at the MMT Observatory, a joint facility of the Smithsonian Institution and the University of Arizona.

The MDM Observatory is operated by Dartmouth College, Columbia University,
Ohio State University, Ohio University, and the University of Michigan.

Some of the data presented in this paper were obtained from the Mikulski Archive for Space Telescopes (MAST) at the Space Telescope Science Institute. The specific observations analyzed can be accessed via \dataset[https://doi.org/10.17909/ew79-we39]{https://doi.org/10.17909/ew79-we39}. STScI is operated by the Association of Universities for Research in Astronomy, Inc., under NASA contract NAS5-26555. Support to MAST for these data is provided by the NASA Office of Space Science via grant NAG5-7584 and by other grants and contracts.

This paper includes data collected with the TESS mission, obtained from the MAST data archive at the Space Telescope Science Institute (STScI). Funding for the TESS mission is provided by the NASA Explorer Program. STScI is operated by the Association of Universities for Research in Astronomy, Inc., under NASA contract NAS 5–26555.

Based on observations obtained with the Samuel Oschin 48-inch Telescope at the Palomar
Observatory as part of the Zwicky Transient Facility project. ZTF is supported by the National
Science Foundation under Grant No. AST-1440341 and a collaboration including Caltech, IPAC,
the Weizmann Institute for Science, the Oskar Klein Center at Stockholm University, the University
of Maryland, the University of Washington, Deutsches Elektronen-Synchrotron and Humboldt
University, Los Alamos National Laboratories, the TANGO Consortium of Taiwan, the University
of Wisconsin at Milwaukee, and Lawrence Berkeley National Laboratories. Operations are
conducted by COO, IPAC, and UW.

This publication makes use of data products from the Two Micron All Sky Survey, which is a joint project of the University of Massachusetts and the Infrared Processing and Analysis Center/California Institute of Technology, funded by the National Aeronautics and Space Administration and the National Science Foundation.

This publication makes use of data products from the Wide-field Infrared Survey Explorer, which is a joint project of the University of California, Los Angeles, and the Jet Propulsion Laboratory/California Institute of Technology, and NEOWISE, which is a project of the Jet Propulsion Laboratory/California Institute of Technology. WISE and NEOWISE are funded by the National Aeronautics and Space Administration.

The authors acknowledge the High Performance Computing Center\footnote{http://www.hpcc.ttu.edu} (HPCC) at Texas Tech University for providing computational resources that have contributed to the research results reported within this paper.

This work made use of Astropy:\footnote{http://www.astropy.org} a community-developed core Python package and an ecosystem of tools and resources for astronomy \citep{astropy2013, astropy2018, astropy2022}. 

\bibliographystyle{aasjournal}

\section{Appendix}
\begin{table}
\center
  \renewcommand{\arraystretch}{1.15} 
  \addtolength{\tabcolsep}{2pt}
	\begin{tabular}{l c c}
    \hline
    \hline
    \multicolumn{1}{C}{\rm Object} &
    \multicolumn{1}{C}{\rm HJD} &
    \multicolumn{1}{C}{$v_{\rm r}$} \\
    \multicolumn{1}{C}{} &
    \multicolumn{1}{C}{($-2450000~{\rm days}$)} &
    \multicolumn{1}{C}{(${\rm km~s^{-1}}$)} \\
    \hline
{J0135$+$2359} & {9192.787768} & {$  133.7\pm  7.0$} \\
{J0135$+$2359} & {9548.737532} & {$    7.4\pm 10.6$} \\
{J0135$+$2359} & {9548.786808} & {$  -17.7\pm 10.0$} \\
{J0135$+$2359} & {9548.834528} & {$  -72.6\pm  9.5$} \\
{J0135$+$2359} & {9549.572076} & {$  110.4\pm 16.5$} \\
{J0135$+$2359} & {9549.712261} & {$  117.2\pm 13.1$} \\
{J0135$+$2359} & {9549.776407} & {$  139.2\pm  9.6$} \\
{J0135$+$2359} & {9549.821702} & {$   74.6\pm 12.7$} \\
{J0135$+$2359} & {9549.827105} & {$   98.2\pm 13.5$} \\
{J0135$+$2359} & {9549.831960} & {$   84.5\pm 14.8$} \\
{J0135$+$2359} & {9549.837437} & {$  103.3\pm 14.2$} \\
{J0135$+$2359} & {9550.562962} & {$  -14.0\pm  8.6$} \\
{J0135$+$2359} & {9550.782942} & {$  113.2\pm  8.5$} \\
{J0135$+$2359} & {9551.560708} & {$ -192.1\pm 10.9$} \\
{J0135$+$2359} & {9551.568185} & {$ -200.6\pm 13.3$} \\
{J0135$+$2359} & {9551.574674} & {$ -171.2\pm  8.7$} \\
{J0135$+$2359} & {9551.664734} & {$ -107.1\pm  8.3$} \\
{J0135$+$2359} & {9551.767718} & {$   -1.4\pm  9.1$} \\
\hline
{J0155$-$4148} & {8778.528269} & {$  120.9\pm  4.5$} \\
{J0155$-$4148} & {8778.747847} & {$   93.1\pm  5.5$} \\
{J0155$-$4148} & {8780.584034} & {$  107.9\pm  9.8$} \\
{J0155$-$4148} & {8780.591087} & {$   62.2\pm  8.2$} \\
{J0155$-$4148} & {8781.538707} & {$  199.1\pm  2.5$} \\
{J0155$-$4148} & {8781.631074} & {$  103.9\pm  2.0$} \\
{\nodata} & {\nodata} & {\nodata} \\
    \hline
    \hline
	\end{tabular}
    \caption{Radial velocity measurements for the 28 new binaries identified in this work. This table has been truncated. The complete version of this table is available in the supplemental data files.}
    \label{table:rv_table}
\end{table}
\newpage
\startlongtable
\begin{deluxetable}{l c c c c c c c}
\tablehead{\colhead{\textsc{source\_id}} & \colhead{R.A.} & \colhead{Decl.} & \colhead{$T_{\rm eff}$} & \colhead{$\log{g}$} & \colhead{Gaia $G$} & \colhead{Gaia ($BP-RP$)} & \colhead{Gaia Parallax} \\ 
\colhead{(Gaia DR3)} & \colhead{(2016.0)} & \colhead{(2016.0)} & \colhead{(K)} & \colhead{(${\rm cm~s^{-2}}$)} & \colhead{(mag)} & \colhead{(mag)} & \colhead{(mas)} } 
\tablecaption{Atmospheric parameters (assuming pure-hydrogen atmospheres) for objects observed as part of our follow-up campaign with $\log{g}>5.0$. Objects which show periodic photometric variability in ZTF DR16 or TESS high-cadence data, and objects which were classified as ELM white dwarf candidates in \citet{pelisoli2019}, are marked. Optical spectroscopy for each object presented in this table is available in a public Zenodo archive \citep{zenodo}.}
\label{table:fig1_table}
\startdata
2882002220454588672 & 00:03:06.730 & $+$40:39:27.119 & $19250\pm470$ & $8.00\pm0.07$ & $18.49$ & $ 0.01\pm0.01$ & $1.71\pm0.32$ \\
2739093475807603584 & 00:03:19.659 & $+$02:26:23.089 & $21790\pm1190$ & $8.35\pm0.18$ & $16.39$ & $-0.098\pm0.005$ & $6.32\pm0.07$ \\
2798386113507604992\footref{footnote:pelisoli} & 00:03:33.944 & $+$20:16:26.238 & $17760\pm540$ & $7.72\pm0.10$ & $18.95$ & $ 0.07\pm0.02$ & $2.25\pm0.25$ \\
2874194833198918400 & 00:03:42.161 & $+$32:44:15.731 & $16410\pm440$ & $7.79\pm0.08$ & $18.22$ & $ 0.22\pm0.01$ & $3.30\pm0.14$ \\
2448821478361661696 & 00:05:04.760 & $-$01:27:08.384 & $31340\pm1380$ & $7.68\pm0.31$ & $16.63$ & $-0.33\pm0.01$ & $3.66\pm0.07$ \\
420251387300980096\footref{footnote:pelisoli} & 00:07:09.001 & $+$53:49:47.474 & $32980\pm1930$ & $6.15\pm0.35$ & $16.17$ & $0.071\pm0.004$ & $0.96\pm0.04$ \\
2859826106009305216\footref{footnote:pelisoli} & 00:12:32.540 & $+$27:47:09.460 & $25990\pm860$ & $7.37\pm0.13$ & $16.98$ & $-0.31\pm0.01$ & $2.89\pm0.09$ \\
2768409887481718272\footref{footnote:pelisoli} & 00:12:45.662 & $+$14:39:56.678 & $14580\pm410$ & $7.38\pm0.08$ & $18.19$ & $ 0.07\pm0.01$ & $2.78\pm0.18$ \\
2800210546895703680\footref{footnote:pelisoli} & 00:15:48.404 & $+$21:27:46.332 & $32970\pm1020$ & $7.35\pm0.21$ & $17.51$ & $-0.30\pm0.01$ & $1.85\pm0.10$ \\
2876543282660374784 & 00:17:44.512 & $+$35:58:26.087 & $18410\pm430$ & $7.95\pm0.07$ & $18.85$ & $ 0.03\pm0.02$ & $1.88\pm0.23$ \\
2863621134816230528\footref{footnote:tess}\footref{footnote:pelisoli} & 00:18:11.384 & $+$33:11:08.826 & $30630\pm590$ & $7.44\pm0.09$ & $18.04$ & $-0.29\pm0.01$ & $1.28\pm0.14$ \\
2855799415254881408 & 00:20:22.382 & $+$26:40:10.956 & $11820\pm220$ & $8.02\pm0.13$ & $16.11$ & $-0.111\pm0.005$ & $10.51\pm0.05$ \\
380560941677424768\footref{footnote:ztf} & 00:33:52.632 & $+$38:55:29.608 & $34130\pm700$ & $7.30\pm0.13$ & $18.34$ & $-0.19\pm0.02$ & $1.19\pm0.17$ \\
4907063048361514496 & 00:40:00.768 & $-$58:40:31.721 & $38240\pm750$ & $7.28\pm0.09$ & $17.58$ & $-0.24\pm0.02$ & $1.14\pm0.08$ \\
2542961560852591744 & 00:40:22.906 & $-$00:21:30.172 & $16470\pm510$ & $8.16\pm0.09$ & $14.85$ & $0.046\pm0.004$ & $18.24\pm0.03$ \\
2550740120286280576 & 00:45:36.928 & $+$02:40:14.491 & $15140\pm480$ & $5.03\pm0.10$ & $18.85$ & $ 0.31\pm0.02$ & $0.93\pm0.27$ \\
2809085018776598528\footref{footnote:pelisoli} & 00:49:11.206 & $+$28:16:02.255 & $27850\pm2870$ & $5.68\pm0.42$ & $16.56$ & $-0.32\pm0.01$ & $0.59\pm0.07$ \\
377520826387065856 & 00:52:04.391 & $+$45:05:33.799 & $11730\pm180$ & $8.21\pm0.11$ & $16.01$ & $-0.039\pm0.004$ & $13.28\pm0.04$ \\
2776836514532338304\footref{footnote:pelisoli} & 00:52:44.411 & $+$13:16:35.972 & $36500\pm1560$ & $7.33\pm0.28$ & $17.94$ & $-0.32\pm0.01$ & $1.11\pm0.13$ \\
374447996328984704\footref{footnote:pelisoli} & 01:01:10.344 & $+$41:06:04.921 & $21740\pm510$ & $7.36\pm0.07$ & $18.41$ & $-0.24\pm0.02$ & $1.68\pm0.17$ \\
2785461702216001152 & 01:07:25.996 & $+$19:09:32.166 & $21940\pm1260$ & $7.56\pm0.19$ & $15.64$ & $-0.276\pm0.005$ & $5.93\pm0.05$ \\
2579742694406440064\footref{footnote:pelisoli} & 01:09:29.505 & $+$09:19:51.845 & $32350\pm1030$ & $7.81\pm0.23$ & $17.37$ & $-0.36\pm0.01$ & $1.53\pm0.11$ \\
2534408386884529920 & 01:12:58.398 & $-$00:59:52.519 & $19380\pm610$ & $7.58\pm0.10$ & $18.84$ & $-0.27\pm0.03$ & $1.09\pm0.25$ \\
373358998783703808 & 01:16:00.832 & $+$42:49:38.323 & $12970\pm200$ & $5.06\pm0.05$ & $18.71$ & $-0.00\pm0.02$ & $0.57\pm0.18$ \\
320040145192523264\footref{footnote:pelisoli} & 01:16:20.951 & $+$33:41:16.094 & $13880\pm240$ & $7.42\pm0.05$ & $18.16$ & $-0.06\pm0.01$ & $2.16\pm0.16$ \\
2482810406432480512\footref{footnote:ztf} & 01:16:51.573 & $-$04:40:47.107 & $45500\pm870$ & $7.58\pm0.10$ & $18.34$ & $-0.25\pm0.02$ & $1.02\pm0.18$ \\
2471069512033475712 & 01:21:49.346 & $-$10:08:00.751 & $17340\pm490$ & $7.95\pm0.09$ & $17.31$ & $ 0.07\pm0.01$ & $8.58\pm0.10$ \\
2563353721975582464\footref{footnote:pelisoli} & 01:25:43.673 & $+$04:55:14.686 & $18120\pm450$ & $7.60\pm0.08$ & $18.15$ & $ 0.19\pm0.01$ & $4.07\pm0.19$ \\
531617690261698304\footref{footnote:tess}\footref{footnote:ztf} & 01:33:38.050 & $+$68:03:32.789 & $34860\pm1030$ & $7.58\pm0.21$ & $16.72$ & $ 0.21\pm0.01$ & $2.60\pm0.05$ \\
296853330452496000 & 01:34:46.322 & $+$28:26:16.202 & $12300\pm330$ & $7.89\pm0.11$ & $16.91$ & $ 0.02\pm0.01$ & $5.68\pm0.07$ \\
315437516503244928 & 01:35:01.880 & $+$31:17:10.212 & $18700\pm500$ & $7.93\pm0.09$ & $19.07$ & $-0.17\pm0.02$ & $2.33\pm0.32$ \\
322765211746880384 & 01:38:52.560 & $+$38:34:01.819 & $34230\pm680$ & $7.83\pm0.12$ & $18.46$ & $-0.36\pm0.02$ & $1.29\pm0.18$ \\
347329027362621056 & 01:40:12.938 & $+$41:29:27.074 & $25380\pm550$ & $7.76\pm0.08$ & $18.36$ & $-0.28\pm0.01$ & $2.13\pm0.20$ \\
319243613441880320 & 01:43:20.592 & $+$36:28:42.978 & $36240\pm1140$ & $8.27\pm0.21$ & $17.23$ & $-0.37\pm0.01$ & $2.39\pm0.09$ \\
5139287360210678144\footref{footnote:pelisoli} & 01:48:11.827 & $-$19:48:14.188 & $15090\pm240$ & $7.74\pm0.05$ & $18.82$ & $ 0.03\pm0.01$ & $1.56\pm0.19$ \\
92185017673400960 & 01:52:27.290 & $+$18:20:14.374 & $34580\pm990$ & $7.81\pm0.20$ & $17.43$ & $-0.19\pm0.01$ & $1.99\pm0.12$ \\
5134761186235143424 & 01:55:58.754 & $-$22:38:22.520 & $23910\pm610$ & $7.58\pm0.09$ & $17.64$ & $-0.32\pm0.01$ & $2.30\pm0.10$ \\
2466766946939761280\footref{footnote:ztf} & 01:58:41.010 & $-$06:28:41.650 & $22740\pm550$ & $7.74\pm0.08$ & $16.91$ & $ 0.12\pm0.01$ & $3.55\pm0.08$ \\
4968866493879269760\footref{footnote:pelisoli} & 01:59:48.100 & $-$35:16:51.229 & $21850\pm720$ & $7.52\pm0.10$ & $17.92$ & $-0.24\pm0.01$ & $1.97\pm0.10$ \\
104648944047154432 & 02:08:47.713 & $+$25:14:08.045 & $20880\pm440$ & $8.05\pm0.06$ & $13.23$ & $-0.208\pm0.004$ & $25.59\pm0.03$ \\
5136945744040876928 & 02:09:30.476 & $-$20:19:33.535 & $27690\pm500$ & $7.53\pm0.07$ & $18.30$ & $-0.29\pm0.02$ & $1.68\pm0.17$ \\
4715247167518325376 & 02:10:35.380 & $-$57:49:57.976 & $26780\pm460$ & $7.58\pm0.06$ & $17.66$ & $-0.33\pm0.01$ & $2.29\pm0.07$ \\
331359166428843904 & 02:13:28.349 & $+$37:16:49.814 & $24300\pm2040$ & $6.79\pm0.28$ & $17.80$ & $-0.29\pm0.01$ & $1.92\pm0.14$ \\
4744277057630500352 & 02:13:54.959 & $-$53:04:53.962 & $26430\pm400$ & $7.42\pm0.05$ & $17.29$ & $-0.33\pm0.01$ & $2.34\pm0.07$ \\
2493466422156808576 & 02:23:17.915 & $-$02:42:52.495 & $28000\pm670$ & $7.36\pm0.10$ & $18.11$ & $-0.05\pm0.02$ & $1.36\pm0.16$ \\
5131327656235273088\footref{footnote:pelisoli} & 02:27:41.362 & $-$18:59:25.656 & $17550\pm2300$ & $7.81\pm0.45$ & $15.98$ & $-0.120\pm0.004$ & $6.45\pm0.04$ \\
76220791737582720 & 02:28:45.867 & $+$15:42:02.599 & $37170\pm1120$ & $7.21\pm0.19$ & $18.26$ & $-0.07\pm0.02$ & $1.20\pm0.19$ \\
131694078232129792 & 02:30:14.426 & $+$30:56:11.627 & $24620\pm550$ & $7.30\pm0.07$ & $18.55$ & $-0.08\pm0.02$ & $1.63\pm0.19$ \\
5130174745278506112 & 02:33:01.035 & $-$20:25:59.635 & $16810\pm590$ & $7.55\pm0.12$ & $19.63$ & $-0.14\pm0.04$ & $1.88\pm0.32$ \\
18117653818752512 & 02:35:47.678 & $+$05:35:24.083 & $24640\pm760$ & $7.31\pm0.11$ & $15.79$ & $-0.36\pm0.01$ & $5.22\pm0.07$ \\
4741227287252719872 & 02:38:43.420 & $-$55:01:27.966 & $18700\pm410$ & $7.41\pm0.07$ & $17.43$ & $-0.22\pm0.01$ & $2.22\pm0.06$ \\
2502595323604529408 & 02:40:09.470 & $+$01:51:49.248 & $17710\pm300$ & $7.41\pm0.05$ & $18.44$ & $-0.19\pm0.02$ & $1.92\pm0.18$ \\
4952731126381731328 & 02:45:09.857 & $-$38:05:40.247 & $14240\pm280$ & $7.84\pm0.05$ & $18.06$ & $-0.09\pm0.01$ & $2.42\pm0.10$ \\
6306661258129280 & 02:48:45.909 & $+$05:39:21.758 & $12410\pm280$ & $7.86\pm0.09$ & $19.04$ & $ 0.13\pm0.02$ & $1.50\pm0.38$ \\
115708721056790144\footref{footnote:pelisoli} & 03:01:45.948 & $+$26:55:44.879 & $30780\pm810$ & $7.76\pm0.16$ & $18.55$ & $-0.08\pm0.02$ & $1.43\pm0.19$ \\
5104166386135159936 & 03:06:39.346 & $-$19:02:24.893 & $35530\pm650$ & $8.20\pm0.09$ & $17.32$ & $-0.08\pm0.01$ & $2.71\pm0.08$ \\
3262674487682440448\footref{footnote:pelisoli} & 03:18:13.332 & $-$01:07:11.806 & $12450\pm1480$ & $7.52\pm0.44$ & $14.71$ & $-0.062\pm0.004$ & $14.25\pm0.03$ \\
3260723782255411456 & 03:23:22.673 & $-$03:54:20.808 & $28440\pm630$ & $7.88\pm0.10$ & $18.50$ & $-0.32\pm0.03$ & $2.22\pm0.24$ \\
237133566847520000\footref{footnote:pelisoli} & 03:38:47.068 & $+$41:34:24.100 & $23460\pm430$ & $5.82\pm0.06$ & $15.09$ & $-0.061\pm0.004$ & $1.65\pm0.03$ \\
69886853143994496\footref{footnote:pelisoli} & 03:44:22.259 & $+$25:14:53.293 & $28280\pm1340$ & $6.45\pm0.22$ & $15.55$ & $-0.006\pm0.004$ & $1.79\pm0.04$ \\
5113854281871286912 & 03:45:32.977 & $-$14:04:58.102 & $11140\pm320$ & $5.02\pm0.10$ & $18.64$ & $ 0.18\pm0.02$ & $0.58\pm0.18$ \\
4858747758657659136 & 03:51:23.934 & $-$34:45:50.306 & $15450\pm450$ & $7.34\pm0.09$ & $18.14$ & $-0.18\pm0.01$ & $2.54\pm0.10$ \\
4887600528615197184 & 03:52:10.364 & $-$30:25:23.981 & $28540\pm890$ & $7.40\pm0.15$ & $17.21$ & $-0.21\pm0.01$ & $2.45\pm0.07$ \\
3251978198050225664 & 04:03:24.704 & $-$03:28:51.614 & $26600\pm450$ & $7.43\pm0.06$ & $16.68$ & $-0.30\pm0.01$ & $3.32\pm0.06$ \\
4844023064578952320\footref{footnote:pelisoli} & 04:04:44.429 & $-$39:50:43.206 & $35120\pm780$ & $7.54\pm0.13$ & $17.66$ & $-0.27\pm0.02$ & $1.58\pm0.07$ \\
4681729620697295360\footref{footnote:tess} & 04:10:33.371 & $-$58:52:03.482 & $32040\pm480$ & $7.28\pm0.06$ & $17.57$ & $-0.15\pm0.02$ & $1.52\pm0.07$ \\
4830937162517646592 & 04:13:45.134 & $-$47:37:25.612 & $18500\pm360$ & $7.84\pm0.06$ & $16.54$ & $-0.14\pm0.01$ & $5.33\pm0.04$ \\
166283580171410816 & 04:20:32.033 & $+$31:09:08.701 & $34450\pm760$ & $7.25\pm0.13$ & $18.70$ & $ 0.05\pm0.02$ & $1.44\pm0.20$ \\
3204352783173513728\footref{footnote:ztf} & 04:31:03.755 & $-$04:17:02.555 & $36010\pm700$ & $7.59\pm0.11$ & $17.92$ & $-0.32\pm0.02$ & $0.98\pm0.14$ \\
3185254609798003328\footref{footnote:pelisoli} & 04:34:38.049 & $-$10:14:07.246 & $17460\pm2920$ & $8.03\pm0.63$ & $16.53$ & $0.004\pm0.004$ & $6.68\pm0.05$ \\
3229440649222191616 & 04:36:59.954 & $-$01:15:53.903 & $31190\pm550$ & $8.05\pm0.08$ & $17.91$ & $-0.30\pm0.01$ & $2.68\pm0.12$ \\
4814354976686021888 & 04:47:19.788 & $-$44:44:10.622 & $20900\pm530$ & $7.69\pm0.08$ & $16.93$ & $-0.27\pm0.01$ & $3.71\pm0.05$ \\
3225359193340812288\footref{footnote:pelisoli} & 04:59:20.296 & $-$02:28:06.650 & $32550\pm1450$ & $6.09\pm0.29$ & $15.05$ & $-0.398\pm0.004$ & $1.11\pm0.04$ \\
2979874526445650560 & 05:01:59.217 & $-$17:09:56.642 & $26160\pm410$ & $7.48\pm0.05$ & $18.42$ & $-0.12\pm0.01$ & $1.30\pm0.14$ \\
3211273483020379136 & 05:07:42.284 & $-$05:50:52.811 & $31150\pm540$ & $7.82\pm0.08$ & $18.31$ & $-0.35\pm0.01$ & $1.50\pm0.17$ \\
2971802603565750016 & 05:31:27.482 & $-$15:56:26.520 & $29490\pm470$ & $7.95\pm0.06$ & $16.55$ & $-0.325\pm0.005$ & $3.97\pm0.06$ \\
4821683294701731456\footref{footnote:pelisoli} & 05:32:25.215 & $-$36:40:57.058 & $24710\pm640$ & $7.30\pm0.09$ & $17.38$ & $-0.30\pm0.01$ & $2.28\pm0.06$ \\
2965034525384116864 & 05:32:46.106 & $-$21:03:40.381 & $27600\pm460$ & $7.53\pm0.06$ & $17.72$ & $-0.22\pm0.01$ & $1.82\pm0.09$ \\
3402486457838073728 & 05:38:02.721 & $+$19:53:02.893 & $37120\pm1080$ & $7.33\pm0.16$ & $18.80$ & $ 0.02\pm0.02$ & $0.96\pm0.29$ \\
480570075502703488\footref{footnote:pelisoli} & 05:39:59.130 & $+$66:30:43.434 & $39310\pm2520$ & $7.13\pm0.27$ & $16.43$ & $-0.228\pm0.005$ & $1.61\pm0.05$ \\
191727718051389312\footref{footnote:pelisoli} & 05:47:06.231 & $+$40:08:22.524 & $28290\pm4040$ & $5.33\pm0.53$ & $15.75$ & $-0.061\pm0.004$ & $1.02\pm0.05$ \\
4805297955811200384\footref{footnote:pelisoli} & 05:47:14.337 & $-$39:22:53.227 & $31790\pm520$ & $7.59\pm0.07$ & $18.26$ & $-0.17\pm0.01$ & $1.25\pm0.11$ \\
2904788774305591168 & 05:49:05.586 & $-$27:51:03.064 & $21370\pm620$ & $7.44\pm0.09$ & $19.18$ & $-0.16\pm0.02$ & $1.34\pm0.18$ \\
3123004865638910208\footref{footnote:pelisoli} & 06:09:30.047 & $+$01:00:43.243 & $35800\pm2690$ & $6.20\pm0.45$ & $16.36$ & $-0.067\pm0.005$ & $0.71\pm0.06$ \\
1106926566694585984 & 06:19:13.539 & $+$69:11:37.003 & $36910\pm750$ & $7.98\pm0.10$ & $16.89$ & $-0.092\pm0.005$ & $3.22\pm0.06$ \\
2999745209821746176\footref{footnote:pelisoli} & 06:20:42.669 & $-$13:07:30.864 & $36350\pm3360$ & $6.15\pm0.53$ & $16.14$ & $-0.170\pm0.005$ & $1.10\pm0.05$ \\
943429565599232128\footref{footnote:pelisoli} & 06:34:49.904 & $+$38:03:51.977 & $22560\pm430$ & $7.18\pm0.06$ & $17.13$ & $-0.19\pm0.01$ & $2.31\pm0.08$ \\
967722411725539456\footref{footnote:pelisoli} & 06:38:40.371 & $+$48:29:50.784 & $30210\pm5070$ & $5.43\pm0.82$ & $16.15$ & $-0.309\pm0.005$ & $0.68\pm0.06$ \\
1141248917368171648\footref{footnote:pelisoli} & 06:43:34.284 & $+$79:20:36.190 & $16930\pm530$ & $7.91\pm0.10$ & $18.86$ & $-0.18\pm0.05$ & $2.24\pm0.16$ \\
3351216303643796096\footref{footnote:pelisoli} & 06:45:32.028 & $+$11:40:03.173 & $27350\pm3840$ & $5.30\pm0.48$ & $16.85$ & $ 0.07\pm0.01$ & $0.67\pm0.06$ \\
3352835983053282304\footref{footnote:pelisoli} & 06:47:34.863 & $+$13:50:11.332 & $31730\pm1640$ & $6.16\pm0.33$ & $16.80$ & $-0.19\pm0.01$ & $0.55\pm0.07$ \\
1106390211177861888\footref{footnote:pelisoli} & 06:47:46.546 & $+$69:15:34.430 & $32030\pm680$ & $7.27\pm0.12$ & $17.90$ & $-0.46\pm0.01$ & $1.47\pm0.11$ \\
1000497070938042496\footref{footnote:pelisoli} & 06:58:37.717 & $+$57:12:35.755 & $34520\pm1960$ & $5.25\pm0.28$ & $16.21$ & $-0.367\pm0.005$ & $0.56\pm0.06$ \\
979204650309974016 & 07:00:04.137 & $+$49:43:38.687 & $35320\pm1580$ & $7.60\pm0.33$ & $17.43$ & $-0.31\pm0.01$ & $2.44\pm0.11$ \\
990086409514581760 & 07:07:17.295 & $+$58:01:36.512 & $23590\pm510$ & $7.42\pm0.07$ & $16.92$ & $-0.30\pm0.01$ & $3.31\pm0.07$ \\
5490015871367837440\footref{footnote:pelisoli} & 07:07:30.007 & $-$56:17:43.206 & $32240\pm480$ & $7.14\pm0.06$ & $15.89$ & $-0.343\pm0.004$ & $2.34\pm0.03$ \\
3161169390978526592\footref{footnote:pelisoli} & 07:12:06.982 & $+$13:23:49.200 & $14790\pm490$ & $7.85\pm0.08$ & $18.83$ & $-0.15\pm0.02$ & $2.00\pm0.23$ \\
1102031472205224832\footref{footnote:ztf} & 07:14:52.278 & $+$66:29:28.648 & $22320\pm760$ & $7.47\pm0.11$ & $18.87$ & $ 0.26\pm0.01$ & $1.39\pm0.17$ \\
989858913686629248 & 07:20:10.814 & $+$57:30:54.979 & $30920\pm1030$ & $7.58\pm0.21$ & $17.87$ & $-0.34\pm0.01$ & $2.17\pm0.13$ \\
900688761191934208 & 07:30:30.416 & $+$41:15:39.827 & $29150\pm800$ & $7.96\pm0.14$ & $17.64$ & $-0.40\pm0.01$ & $2.66\pm0.12$ \\
3169165486210495488\footref{footnote:pelisoli} & 07:34:29.881 & $+$16:48:22.968 & $16540\pm560$ & $8.02\pm0.10$ & $19.02$ & $ 0.04\pm0.04$ & $2.04\pm0.30$ \\
3165507445385638528 & 07:42:11.606 & $+$14:53:17.970 & $17080\pm510$ & $7.74\pm0.09$ & $19.02$ & $-0.01\pm0.03$ & $2.32\pm0.24$ \\
671699491725392128\footref{footnote:pelisoli} & 07:45:44.371 & $+$18:58:49.631 & $15620\pm400$ & $7.69\pm0.08$ & $17.54$ & $ 0.07\pm0.01$ & $3.93\pm0.10$ \\
668912126670207360\footref{footnote:pelisoli} & 08:02:42.951 & $+$17:50:00.848 & $21930\pm1520$ & $7.66\pm0.23$ & $16.27$ & $-0.216\pm0.005$ & $4.23\pm0.06$ \\
3063885736023355904\footref{footnote:pelisoli} & 08:06:50.022 & $-$07:16:36.109 & $13640\pm550$ & $6.46\pm0.14$ & $18.55$ & $-0.07\pm0.02$ & $0.93\pm0.18$ \\
653228761532233216\footref{footnote:pelisoli} & 08:07:39.280 & $+$13:21:11.434 & $15560\pm1910$ & $7.44\pm0.43$ & $16.43$ & $0.097\pm0.004$ & $8.71\pm0.05$ \\
3096525945581800576\footref{footnote:ztf} & 08:11:26.652 & $+$05:39:11.441 & $26860\pm720$ & $7.87\pm0.11$ & $16.54$ & $ 0.02\pm0.01$ & $4.57\pm0.06$ \\
3070115057805517568\footref{footnote:pelisoli} & 08:18:41.061 & $-$02:14:13.434 & $33450\pm1740$ & $5.13\pm0.33$ & $16.49$ & $-0.35\pm0.01$ & $0.50\pm0.06$ \\
3090361426264736000\footref{footnote:pelisoli} & 08:20:00.642 & $+$02:23:27.107 & $14760\pm370$ & $7.53\pm0.07$ & $18.44$ & $-0.01\pm0.02$ & $2.29\pm0.19$ \\
928585329693880832\footref{footnote:pelisoli} & 08:20:10.339 & $+$45:43:01.704 & $17360\pm250$ & $7.46\pm0.04$ & $18.21$ & $-0.08\pm0.01$ & $2.38\pm0.18$ \\
3090457599171887104 & 08:21:01.664 & $+$02:32:09.316 & $23240\pm950$ & $7.54\pm0.14$ & $18.81$ & $ 0.35\pm0.02$ & $1.40\pm0.22$ \\
1123423639154702464\footref{footnote:pelisoli} & 08:21:42.669 & $+$73:55:29.406 & $35660\pm1370$ & $6.06\pm0.24$ & $14.43$ & $-0.452\pm0.004$ & $1.75\pm0.04$ \\
709064710767864064\footref{footnote:pelisoli} & 08:23:59.152 & $+$31:11:51.893 & $19680\pm390$ & $7.57\pm0.06$ & $18.29$ & $-0.20\pm0.01$ & $1.98\pm0.18$ \\
5196549827003398656 & 08:28:48.624 & $-$80:19:35.026 & $24480\pm370$ & $7.54\pm0.05$ & $15.33$ & $-0.328\pm0.004$ & $7.14\pm0.03$ \\
707909910025230080 & 08:28:55.281 & $+$29:21:00.310 & $15560\pm390$ & $7.74\pm0.07$ & $18.85$ & $-0.00\pm0.03$ & $1.95\pm0.23$ \\
3073641569552516864\footref{footnote:pelisoli} & 08:36:01.849 & $-$00:44:22.974 & $26820\pm480$ & $6.02\pm0.06$ & $15.92$ & $-0.349\pm0.004$ & $1.06\pm0.06$ \\
662771594747150976 & 08:36:12.031 & $+$19:17:55.853 & $33230\pm2400$ & $5.36\pm0.48$ & $15.69$ & $-0.092\pm0.005$ & $0.53\pm0.07$ \\
664631178147534720\footref{footnote:pelisoli} & 08:36:19.698 & $+$20:57:47.358 & $30820\pm1540$ & $6.03\pm0.29$ & $16.31$ & $-0.43\pm0.01$ & $0.54\pm0.09$ \\
3078353045597053696\footref{footnote:pelisoli} & 08:40:14.857 & $+$00:42:39.028 & $33580\pm1900$ & $5.92\pm0.37$ & $16.47$ & $-0.325\pm0.005$ & $0.59\pm0.05$ \\
582675157663938304 & 08:48:23.744 & $+$05:08:44.930 & $16100\pm330$ & $7.78\pm0.07$ & $17.42$ & $-0.12\pm0.01$ & $2.59\pm0.09$ \\
605052074715179008\footref{footnote:pelisoli} & 08:54:08.733 & $+$12:22:26.504 & $15020\pm280$ & $7.50\pm0.05$ & $18.50$ & $-0.02\pm0.01$ & $2.41\pm0.20$ \\
684882528766947456 & 08:57:45.557 & $+$20:38:50.381 & $11400\pm310$ & $5.04\pm0.10$ & $18.20$ & $ 0.22\pm0.01$ & $0.90\pm0.18$ \\
718805520500404608 & 09:00:14.503 & $+$37:57:49.241 & $22340\pm540$ & $7.48\pm0.08$ & $19.15$ & $-0.38\pm0.03$ & $0.95\pm0.27$ \\
611215077906912640\footref{footnote:pelisoli} & 09:03:00.839 & $+$17:21:32.170 & $11690\pm210$ & $7.83\pm0.06$ & $18.38$ & $ 0.01\pm0.01$ & $2.50\pm0.18$ \\
1009623532844468480 & 09:08:46.276 & $+$45:16:52.561 & $11870\pm330$ & $7.87\pm0.10$ & $18.27$ & $-0.09\pm0.01$ & $2.94\pm0.16$ \\
1009618997359086848 & 09:09:19.116 & $+$45:10:51.697 & $16700\pm460$ & $8.00\pm0.08$ & $19.04$ & $-0.04\pm0.02$ & $2.14\pm0.30$ \\
638506468336791296 & 09:17:29.690 & $+$21:43:33.452 & $20540\pm540$ & $7.85\pm0.08$ & $18.85$ & $-0.02\pm0.02$ & $1.51\pm0.29$ \\
3845945232756311936\footref{footnote:tess}\footref{footnote:pelisoli} & 09:23:37.323 & $+$03:53:09.208 & $31520\pm730$ & $7.35\pm0.14$ & $17.53$ & $-0.35\pm0.01$ & $1.61\pm0.12$ \\
817452604232788224 & 09:24:39.388 & $+$42:36:33.415 & $11260\pm260$ & $7.87\pm0.08$ & $18.79$ & $-0.15\pm0.02$ & $1.41\pm0.18$ \\
5689579636989620608\footref{footnote:pelisoli} & 09:25:09.586 & $-$14:11:22.556 & $26680\pm3340$ & $7.07\pm0.48$ & $15.39$ & $-0.374\pm0.004$ & $4.07\pm0.04$ \\
1063375422215251456 & 09:36:36.353 & $+$62:56:05.593 & $15550\pm580$ & $7.54\pm0.12$ & $18.24$ & $ 0.55\pm0.01$ & $2.54\pm0.12$ \\
794276587043539456\footref{footnote:pelisoli} & 09:37:08.551 & $+$33:34:04.141 & $24620\pm830$ & $7.33\pm0.12$ & $16.36$ & $-0.323\pm0.005$ & $3.11\pm0.08$ \\
616823209979081472\footref{footnote:ztf} & 09:52:50.473 & $+$15:53:04.013 & $22810\pm620$ & $7.58\pm0.09$ & $18.31$ & $ 0.42\pm0.02$ & $1.35\pm0.20$ \\
3879591353717543040 & 09:53:02.996 & $+$10:32:20.094 & $11610\pm240$ & $7.81\pm0.07$ & $18.79$ & $-0.13\pm0.02$ & $1.73\pm0.25$ \\
1146182357322101120 & 10:07:52.208 & $+$81:40:46.549 & $19280\pm740$ & $7.66\pm0.13$ & $18.63$ & $ 0.30\pm0.01$ & $1.41\pm0.21$ \\
3752200596493839232\footref{footnote:pelisoli} & 10:24:32.206 & $-$14:34:20.503 & $ 9290\pm150$ & $8.20\pm0.10$ & $16.81$ & $-0.13\pm0.01$ & $2.82\pm0.09$ \\
3767870664414250112\footref{footnote:ztf} & 10:26:53.472 & $-$10:13:30.194 & $22150\pm670$ & $7.53\pm0.09$ & $18.35$ & $-0.15\pm0.03$ & $1.65\pm0.19$ \\
5467851842959399808\footref{footnote:tess}\footref{footnote:ztf} & 10:30:39.621 & $-$27:54:38.876 & $32020\pm800$ & $7.75\pm0.16$ & $18.31$ & $-0.04\pm0.02$ & $1.31\pm0.18$ \\
1048058916002814080 & 10:33:50.070 & $+$60:44:01.421 & $19570\pm530$ & $7.50\pm0.09$ & $18.54$ & $-0.24\pm0.01$ & $1.86\pm0.14$ \\
5468388683808323328 & 10:35:51.946 & $-$26:43:01.996 & $23340\pm810$ & $7.48\pm0.12$ & $18.92$ & $-0.27\pm0.02$ & $1.37\pm0.24$ \\
1071978142925329280 & 10:36:05.668 & $+$67:38:17.232 & $ 9370\pm160$ & $8.37\pm0.16$ & $17.67$ & $-0.26\pm0.01$ & $2.22\pm0.08$ \\
3754558774057309568 & 10:36:42.062 & $-$11:47:42.065 & $23940\pm590$ & $7.86\pm0.09$ & $19.56$ & $-0.25\pm0.04$ & $1.26\pm0.37$ \\
3777010874511580800\footref{footnote:pelisoli} & 10:51:15.599 & $-$04:54:52.938 & $21230\pm610$ & $7.39\pm0.09$ & $17.92$ & $-0.21\pm0.01$ & $2.14\pm0.16$ \\
3801572246288990720 & 10:51:59.289 & $-$02:48:05.868 & $27700\pm460$ & $7.29\pm0.06$ & $17.49$ & $-0.34\pm0.01$ & $1.26\pm0.20$ \\
3865904461176307840\footref{footnote:pelisoli} & 10:52:13.140 & $+$08:06:31.482 & $14480\pm660$ & $7.50\pm0.15$ & $17.73$ & $ 0.06\pm0.01$ & $3.31\pm0.13$ \\
3758426443647179008 & 11:06:21.270 & $-$10:56:47.407 & $15870\pm420$ & $7.70\pm0.09$ & $18.17$ & $-0.11\pm0.01$ & $1.98\pm0.16$ \\
3803538104360099328 & 11:06:27.670 & $-$01:05:14.698 & $30280\pm630$ & $7.56\pm0.11$ & $16.47$ & $ 0.06\pm0.01$ & $3.03\pm0.08$ \\
3783864852042143616 & 11:22:56.309 & $-$06:45:05.792 & $25160\pm1180$ & $7.25\pm0.18$ & $17.97$ & $ 0.15\pm0.01$ & $1.77\pm0.13$ \\
3812632092018399104\footref{footnote:pelisoli} & 11:24:21.658 & $+$03:30:26.003 & $20360\pm500$ & $7.65\pm0.07$ & $18.16$ & $-0.23\pm0.02$ & $2.02\pm0.16$ \\
3796545519645331584 & 11:27:21.271 & $-$02:08:37.464 & $23520\pm720$ & $7.90\pm0.10$ & $16.39$ & $-0.260\pm0.005$ & $5.39\pm0.07$ \\
3484922002822079872\footref{footnote:pelisoli} & 11:35:24.140 & $-$26:54:41.544 & $18160\pm350$ & $7.52\pm0.06$ & $17.62$ & $-0.12\pm0.01$ & $2.91\pm0.11$ \\
3895861965440903680\footref{footnote:pelisoli} & 11:51:24.105 & $+$03:41:47.263 & $19950\pm470$ & $7.61\pm0.07$ & $18.13$ & $-0.24\pm0.01$ & $2.14\pm0.14$ \\
3892319850076956032\footref{footnote:pelisoli} & 11:53:16.969 & $+$01:45:08.284 & $15000\pm620$ & $7.46\pm0.11$ & $18.74$ & $ 0.01\pm0.01$ & $2.15\pm0.26$ \\
4013124296427540096 & 12:08:34.821 & $+$29:08:50.208 & $15980\pm430$ & $7.82\pm0.07$ & $18.89$ & $-0.19\pm0.02$ & $2.06\pm0.30$ \\
3501205731372521984 & 12:35:49.147 & $-$23:28:06.280 & $19410\pm610$ & $7.73\pm0.11$ & $18.98$ & $-0.02\pm0.02$ & $1.43\pm0.24$ \\
6157479215371911040 & 12:38:33.842 & $-$35:07:23.412 & $16130\pm320$ & $7.45\pm0.06$ & $19.00$ & $-0.09\pm0.02$ & $1.30\pm0.22$ \\
3676893915531705984 & 12:47:29.573 & $-$05:39:23.036 & $25440\pm530$ & $7.69\pm0.07$ & $18.44$ & $-0.26\pm0.02$ & $1.31\pm0.19$ \\
3525671553013347968\footref{footnote:pelisoli} & 12:51:41.084 & $-$13:50:12.520 & $15440\pm390$ & $7.54\pm0.08$ & $18.52$ & $ 0.16\pm0.02$ & $2.62\pm0.22$ \\
3625674673885031680\footref{footnote:pelisoli} & 12:59:29.143 & $-$10:17:48.257 & $16820\pm410$ & $7.58\pm0.08$ & $18.14$ & $ 0.12\pm0.01$ & $2.90\pm0.16$ \\
6193688224923108480 & 13:13:23.141 & $-$25:03:08.309 & $28540\pm440$ & $7.44\pm0.05$ & $18.58$ & $-0.25\pm0.02$ & $1.28\pm0.19$ \\
1446939125851716864 & 13:16:26.562 & $+$24:24:50.440 & $22670\pm1380$ & $7.28\pm0.20$ & $17.94$ & $-0.32\pm0.01$ & $2.05\pm0.13$ \\
3636151129911425408 & 13:19:09.855 & $-$04:13:14.275 & $16100\pm510$ & $7.68\pm0.09$ & $17.41$ & $-0.01\pm0.01$ & $2.42\pm0.29$ \\
1718367456798754176 & 13:21:30.497 & $+$79:58:32.311 & $27590\pm700$ & $8.00\pm0.12$ & $18.52$ & $-0.09\pm0.02$ & $1.93\pm0.18$ \\
3631612758228909696 & 13:29:04.203 & $-$06:54:00.000 & $15980\pm460$ & $8.22\pm0.09$ & $19.58$ & $-0.19\pm0.03$ & $1.85\pm0.38$ \\
3741493212960873472 & 13:34:28.970 & $+$14:06:47.412 & $16200\pm690$ & $7.80\pm0.16$ & $18.88$ & $-0.15\pm0.02$ & $1.59\pm0.23$ \\
6293400838501106432 & 13:45:45.625 & $-$18:16:46.675 & $18630\pm490$ & $7.49\pm0.08$ & $19.22$ & $-0.07\pm0.03$ & $1.42\pm0.28$ \\
3725586917543559936 & 13:46:08.455 & $+$10:27:09.882 & $20280\pm880$ & $7.69\pm0.16$ & $19.12$ & $-0.11\pm0.02$ & $1.27\pm0.28$ \\
3728966476985207936\footref{footnote:pelisoli} & 13:51:53.790 & $+$14:09:45.299 & $18690\pm1110$ & $8.16\pm0.21$ & $15.25$ & $-0.190\pm0.004$ & $8.64\pm0.03$ \\
6121727078670155648 & 14:02:16.856 & $-$35:26:26.632 & $21500\pm590$ & $7.51\pm0.08$ & $19.22$ & $ 0.24\pm0.04$ & $1.35\pm0.38$ \\
6302470194522928384\footref{footnote:pelisoli} & 14:04:55.459 & $-$13:01:43.558 & $16030\pm370$ & $8.01\pm0.07$ & $18.54$ & $-0.11\pm0.02$ & $2.07\pm0.21$ \\
3640987851498162048 & 14:12:32.008 & $-$05:49:43.979 & $14050\pm620$ & $7.45\pm0.10$ & $18.37$ & $-0.10\pm0.02$ & $1.81\pm0.19$ \\
6216651490910555008\footref{footnote:pelisoli} & 14:32:14.140 & $-$31:32:06.803 & $35310\pm810$ & $7.35\pm0.14$ & $18.18$ & $-0.08\pm0.02$ & $1.31\pm0.17$ \\
1603554764703627520\footref{footnote:pelisoli} & 14:36:33.304 & $+$50:10:26.843 & $17010\pm270$ & $6.68\pm0.05$ & $18.42$ & $-0.19\pm0.01$ & $1.05\pm0.13$ \\
6323667889648971520 & 14:45:12.622 & $-$12:34:24.856 & $38300\pm1940$ & $7.62\pm0.28$ & $16.83$ & $-0.34\pm0.01$ & $3.40\pm0.10$ \\
6338539467313753856\footref{footnote:ztf} & 14:45:32.863 & $-$04:05:32.788 & $19990\pm800$ & $7.61\pm0.14$ & $18.39$ & $-0.09\pm0.01$ & $2.06\pm0.26$ \\
1185747286815299584 & 14:46:06.934 & $+$14:24:20.959 & $16700\pm480$ & $7.56\pm0.08$ & $18.91$ & $-0.01\pm0.02$ & $1.69\pm0.24$ \\
1171774040214623744\footref{footnote:pelisoli} & 14:46:16.111 & $+$07:35:53.934 & $11180\pm300$ & $7.82\pm0.15$ & $18.43$ & $-0.20\pm0.02$ & $1.84\pm0.21$ \\
6202134673248731648 & 14:55:21.274 & $-$36:09:18.914 & $13220\pm380$ & $5.51\pm0.10$ & $17.94$ & $-0.05\pm0.01$ & $0.88\pm0.14$ \\
1587468291113937152\footref{footnote:pelisoli} & 14:58:00.877 & $+$47:15:50.616 & $15800\pm530$ & $7.56\pm0.11$ & $18.86$ & $ 0.04\pm0.02$ & $2.04\pm0.17$ \\
6320134639096976512 & 15:07:12.554 & $-$09:11:09.085 & $17930\pm440$ & $7.55\pm0.08$ & $18.62$ & $-0.04\pm0.02$ & $1.75\pm0.20$ \\
6334550507847433856 & 15:12:45.575 & $-$05:57:16.974 & $25180\pm900$ & $7.55\pm0.15$ & $18.46$ & $-0.15\pm0.02$ & $2.05\pm0.19$ \\
1290428254138905216 & 15:13:50.075 & $+$33:44:26.059 & $15070\pm470$ & $7.69\pm0.09$ & $17.29$ & $-0.12\pm0.01$ & $3.68\pm0.06$ \\
6209808611575247104 & 15:29:55.238 & $-$29:22:11.068 & $30980\pm1040$ & $7.47\pm0.21$ & $18.44$ & $ 0.07\pm0.01$ & $1.67\pm0.20$ \\
1272514216129754496 & 15:36:41.442 & $+$28:43:49.188 & $22580\pm690$ & $7.59\pm0.10$ & $18.38$ & $-0.26\pm0.01$ & $1.89\pm0.13$ \\
1190410040751247488 & 15:38:46.021 & $+$12:07:47.874 & $13170\pm430$ & $7.88\pm0.08$ & $18.53$ & $ 0.03\pm0.02$ & $2.34\pm0.21$ \\
1209860985561813248 & 15:39:09.221 & $+$18:48:17.863 & $12610\pm320$ & $5.18\pm0.08$ & $17.96$ & $ 0.22\pm0.01$ & $1.24\pm0.33$ \\
6260329800780772608\footref{footnote:ztf}\footref{footnote:pelisoli} & 15:44:25.902 & $-$18:16:51.305 & $35190\pm820$ & $7.13\pm0.15$ & $17.52$ & $-0.20\pm0.02$ & $1.00\pm0.11$ \\
1646202037605812352\footref{footnote:ztf} & 15:46:48.400 & $+$67:40:50.549 & $27840\pm640$ & $7.62\pm0.11$ & $18.04$ & $-0.01\pm0.01$ & $1.60\pm0.09$ \\
6243194942858258304\footref{footnote:pelisoli} & 16:01:26.624 & $-$22:44:47.033 & $30280\pm900$ & $7.37\pm0.17$ & $18.27$ & $-0.16\pm0.01$ & $1.75\pm0.17$ \\
4356545788311602944 & 16:13:30.572 & $-$04:24:15.199 & $28920\pm480$ & $7.24\pm0.06$ & $17.47$ & $-0.16\pm0.01$ & $1.70\pm0.12$ \\
4450336810219872896 & 16:14:04.210 & $+$07:03:18.328 & $17200\pm560$ & $7.78\pm0.12$ & $19.03$ & $-0.08\pm0.02$ & $1.20\pm0.26$ \\
1623581922326511104 & 16:23:23.508 & $+$58:03:53.597 & $25300\pm800$ & $7.62\pm0.14$ & $19.04$ & $-0.37\pm0.02$ & $0.90\pm0.17$ \\
1411455519097674368 & 16:32:42.394 & $+$49:36:14.602 & $12400\pm220$ & $5.54\pm0.06$ & $18.04$ & $ 0.16\pm0.01$ & $0.75\pm0.09$ \\
1426734882432542464\footref{footnote:pelisoli} & 16:39:07.170 & $+$54:17:47.015 & $12740\pm330$ & $7.84\pm0.10$ & $18.17$ & $ 0.02\pm0.02$ & $2.88\pm0.11$ \\
4333980859071542016 & 16:43:57.872 & $-$11:52:37.772 & $32220\pm1260$ & $6.21\pm0.25$ & $18.28$ & $ 0.06\pm0.01$ & $0.57\pm0.17$ \\
1705869136327269632 & 16:47:55.848 & $+$78:02:22.207 & $21830\pm670$ & $7.74\pm0.10$ & $16.96$ & $-0.20\pm0.01$ & $3.50\pm0.06$ \\
1407607121025145088 & 16:52:48.734 & $+$46:34:56.996 & $25110\pm740$ & $7.47\pm0.10$ & $18.33$ & $-0.23\pm0.02$ & $1.65\pm0.12$ \\
1408695878054997376\footref{footnote:pelisoli} & 16:52:54.305 & $+$48:04:05.862 & $14960\pm420$ & $7.41\pm0.08$ & $18.79$ & $-0.12\pm0.02$ & $1.95\pm0.15$ \\
4545581528237665024 & 16:56:34.296 & $+$14:21:23.486 & $12760\pm500$ & $8.29\pm0.09$ & $19.34$ & $ 0.22\pm0.02$ & $1.83\pm0.34$ \\
1337970174853051392\footref{footnote:ztf} & 17:01:25.279 & $+$34:35:30.260 & $17750\pm490$ & $7.58\pm0.09$ & $18.94$ & $-0.03\pm0.02$ & $1.35\pm0.16$ \\
4568269229124267904 & 17:08:16.358 & $+$22:25:51.074 & $21500\pm350$ & $6.91\pm0.05$ & $19.48$ & $-0.15\pm0.02$ & $0.93\pm0.26$ \\
1420761029600606592\footref{footnote:tess}\footref{footnote:ztf} & 17:24:06.145 & $+$56:20:03.073 & $39530\pm4180$ & $7.84\pm0.35$ & $16.26$ & $-0.10\pm0.01$ & $2.48\pm0.04$ \\
4601834943998509440 & 17:29:16.269 & $+$32:08:40.438 & $12040\pm590$ & $8.09\pm0.21$ & $19.18$ & $-0.11\pm0.02$ & $1.45\pm0.19$ \\
1335938792758457088 & 17:31:56.687 & $+$34:56:18.726 & $32860\pm810$ & $8.15\pm0.15$ & $18.48$ & $-0.30\pm0.02$ & $1.68\pm0.13$ \\
5921549652415561856 & 17:35:44.646 & $-$54:01:52.511 & $21670\pm760$ & $4.99\pm0.10$ & $18.28$ & $-0.10\pm0.01$ & $0.62\pm0.17$ \\
1349206114240568960 & 17:37:49.760 & $+$43:33:28.476 & $24040\pm1370$ & $7.36\pm0.20$ & $19.86$ & $ 0.25\pm0.04$ & $2.01\pm0.27$ \\
4607711764930567296 & 17:48:33.469 & $+$35:03:56.747 & $23100\pm920$ & $7.88\pm0.13$ & $18.40$ & $-0.03\pm0.02$ & $1.65\pm0.12$ \\
4580948105722037248 & 17:56:10.500 & $+$24:33:58.198 & $18810\pm620$ & $7.24\pm0.11$ & $17.78$ & $-0.16\pm0.01$ & $2.63\pm0.09$ \\
4580931338170148096 & 17:58:22.157 & $+$24:46:45.725 & $21210\pm1640$ & $7.51\pm0.28$ & $19.52$ & $-0.03\pm0.03$ & $1.32\pm0.27$ \\
4583332435345090432 & 18:01:12.127 & $+$27:05:19.187 & $26040\pm1380$ & $7.15\pm0.21$ & $15.46$ & $-0.300\pm0.004$ & $6.14\pm0.03$ \\
2122667052087422848\footref{footnote:pelisoli} & 18:04:27.825 & $+$48:18:39.539 & $18040\pm510$ & $7.66\pm0.09$ & $18.96$ & $-0.12\pm0.02$ & $1.51\pm0.15$ \\
6652309883673792256 & 18:08:21.306 & $-$55:24:57.704 & $29940\pm460$ & $7.40\pm0.05$ & $17.65$ & $-0.13\pm0.01$ & $1.78\pm0.09$ \\
6417682795318622592\footref{footnote:pelisoli} & 18:09:53.230 & $-$74:43:37.769 & $27250\pm1070$ & $7.08\pm0.17$ & $18.33$ & $ 0.02\pm0.01$ & $1.52\pm0.11$ \\
2115751643640408960\footref{footnote:ztf} & 18:11:43.485 & $+$44:55:01.099 & $27050\pm570$ & $7.38\pm0.09$ & $17.37$ & $-0.08\pm0.01$ & $2.21\pm0.06$ \\
2123798209034064512\footref{footnote:ztf} & 18:17:00.609 & $+$50:00:50.008 & $32830\pm660$ & $7.82\pm0.12$ & $17.67$ & $ 0.20\pm0.01$ & $1.60\pm0.07$ \\
4509763077242737152\footref{footnote:pelisoli} & 18:27:51.851 & $+$15:16:35.663 & $13040\pm430$ & $7.95\pm0.09$ & $19.29$ & $ 0.23\pm0.02$ & $1.84\pm0.31$ \\
6650799738811684992 & 18:35:31.234 & $-$54:55:55.434 & $24890\pm440$ & $7.29\pm0.06$ & $18.35$ & $-0.15\pm0.02$ & $1.69\pm0.17$ \\
6636129539276532608\footref{footnote:pelisoli} & 18:36:23.992 & $-$59:51:31.291 & $30610\pm710$ & $8.06\pm0.14$ & $18.09$ & $ 0.09\pm0.01$ & $2.00\pm0.14$ \\
2146494091832635520\footref{footnote:tess}\footref{footnote:ztf} & 18:47:16.639 & $+$53:37:49.998 & $34070\pm770$ & $7.82\pm0.14$ & $17.63$ & $-0.06\pm0.01$ & $1.29\pm0.07$ \\
6651387427778240128 & 18:49:41.321 & $-$53:02:49.114 & $30890\pm720$ & $7.14\pm0.13$ & $17.94$ & $-0.11\pm0.02$ & $1.13\pm0.13$ \\
2252265701675503616 & 19:06:00.874 & $+$62:39:23.713 & $13570\pm210$ & $5.34\pm0.05$ & $17.96$ & $-0.06\pm0.01$ & $0.51\pm0.10$ \\
6744390099848377472 & 19:32:54.642 & $-$34:04:16.882 & $23610\pm980$ & $7.40\pm0.13$ & $17.77$ & $-0.10\pm0.01$ & $2.69\pm0.13$ \\
6645284902019884928 & 19:36:39.332 & $-$52:46:00.102 & $30770\pm1790$ & $8.06\pm0.41$ & $17.97$ & $-0.09\pm0.02$ & $1.55\pm0.14$ \\
6864866513780069120 & 19:49:46.727 & $-$21:44:11.526 & $20850\pm400$ & $7.62\pm0.06$ & $17.51$ & $-0.12\pm0.01$ & $2.58\pm0.10$ \\
6447432350150267392 & 19:53:14.631 & $-$58:02:22.816 & $34200\pm600$ & $7.91\pm0.09$ & $17.05$ & $ 0.21\pm0.01$ & $2.92\pm0.07$ \\
6426888701236987392\footref{footnote:tess} & 19:53:37.608 & $-$67:44:23.384 & $36690\pm1730$ & $7.72\pm0.32$ & $16.82$ & $-0.06\pm0.01$ & $2.11\pm0.05$ \\
6851161273137814528\footref{footnote:ztf}\footref{footnote:pelisoli} & 19:59:01.117 & $-$25:11:43.080 & $32030\pm560$ & $7.67\pm0.09$ & $17.49$ & $ 0.01\pm0.01$ & $2.35\pm0.10$ \\
6867085976780414848\footref{footnote:pelisoli} & 20:00:28.593 & $-$19:30:50.785 & $31860\pm3680$ & $6.52\pm0.73$ & $17.96$ & $-0.09\pm0.01$ & $1.51\pm0.15$ \\
6874698381114836352\footref{footnote:pelisoli} & 20:12:26.489 & $-$15:03:38.736 & $27200\pm520$ & $7.42\pm0.07$ & $18.25$ & $-0.12\pm0.01$ & $1.77\pm0.17$ \\
6904576475726842240 & 20:26:41.354 & $-$09:42:03.582 & $38900\pm3530$ & $7.51\pm0.44$ & $16.39$ & $-0.19\pm0.01$ & $4.27\pm0.07$ \\
1812357593394310400\footref{footnote:pelisoli} & 20:39:43.468 & $+$18:09:34.877 & $31370\pm2450$ & $6.22\pm0.45$ & $15.84$ & $-0.290\pm0.004$ & $0.86\pm0.04$ \\
6677077521256040576\footref{footnote:pelisoli} & 20:50:18.091 & $-$42:19:07.151 & $21770\pm360$ & $7.58\pm0.05$ & $17.85$ & $-0.26\pm0.01$ & $1.13\pm0.14$ \\
1757877204553173248\footref{footnote:pelisoli} & 20:52:36.182 & $+$12:06:43.542 & $33130\pm1070$ & $6.45\pm0.21$ & $16.39$ & $-0.257\pm0.005$ & $0.71\pm0.07$ \\
6805523324308751232 & 20:53:52.091 & $-$25:35:15.475 & $24270\pm830$ & $7.30\pm0.12$ & $18.19$ & $-0.28\pm0.01$ & $1.43\pm0.19$ \\
1734921287254144640\footref{footnote:pelisoli} & 20:56:12.309 & $+$05:25:14.866 & $32410\pm3460$ & $5.49\pm0.63$ & $16.34$ & $-0.343\pm0.005$ & $0.66\pm0.06$ \\
1764456441613885952 & 20:57:30.511 & $+$16:50:14.611 & $22310\pm1250$ & $7.71\pm0.19$ & $16.56$ & $-0.217\pm0.005$ & $4.76\pm0.05$ \\
6479867638227626240\footref{footnote:pelisoli} & 21:02:35.668 & $-$48:20:01.720 & $21750\pm550$ & $7.59\pm0.08$ & $17.21$ & $-0.19\pm0.01$ & $3.08\pm0.09$ \\
6787236251212992896 & 21:08:40.254 & $-$31:32:17.113 & $17270\pm260$ & $7.99\pm0.05$ & $17.54$ & $-0.12\pm0.01$ & $1.98\pm0.25$ \\
6912586070040308096\footref{footnote:pelisoli} & 21:10:35.318 & $-$03:58:25.378 & $10450\pm150$ & $7.81\pm0.06$ & $18.93$ & $ 0.18\pm0.02$ & $2.60\pm0.25$ \\
6579921027395679104 & 21:13:54.811 & $-$43:48:22.496 & $19210\pm470$ & $8.13\pm0.08$ & $18.64$ & $-0.25\pm0.02$ & $1.63\pm0.21$ \\
1847044402111714944 & 21:27:36.602 & $+$26:05:43.559 & $15040\pm470$ & $7.52\pm0.10$ & $19.44$ & $ 0.13\pm0.03$ & $1.85\pm0.27$ \\
1745828481256730368\footref{footnote:pelisoli} & 21:28:23.751 & $+$11:44:51.972 & $34930\pm3530$ & $5.84\pm0.62$ & $15.77$ & $-0.361\pm0.004$ & $0.94\pm0.05$ \\
6816044860312027264 & 21:30:07.548 & $-$22:59:39.732 & $16850\pm560$ & $7.64\pm0.11$ & $18.66$ & $-0.10\pm0.01$ & $1.57\pm0.27$ \\
6828182472250027776\footref{footnote:pelisoli} & 21:30:33.081 & $-$22:18:02.876 & $34170\pm1290$ & $8.01\pm0.29$ & $16.30$ & $-0.374\pm0.005$ & $1.79\pm0.07$ \\
1797040331265541376\footref{footnote:pelisoli} & 21:32:38.897 & $+$22:54:47.650 & $33290\pm1970$ & $7.13\pm0.42$ & $15.67$ & $-0.234\pm0.004$ & $1.05\pm0.04$ \\
6814484584593326720\footref{footnote:ztf} & 21:34:33.314 & $-$25:00:05.080 & $32620\pm580$ & $7.64\pm0.09$ & $17.53$ & $-0.03\pm0.01$ & $1.62\pm0.11$ \\
6458294666038347520 & 21:34:48.160 & $-$59:11:51.709 & $34950\pm650$ & $7.87\pm0.11$ & $18.14$ & $-0.51\pm0.01$ & $1.91\pm0.13$ \\
2672992211134257152\footref{footnote:ztf} & 21:44:54.893 & $-$04:09:18.277 & $28720\pm630$ & $7.61\pm0.09$ & $17.79$ & $ 0.11\pm0.01$ & $2.34\pm0.15$ \\
1769307791858602880 & 21:49:11.106 & $+$15:06:37.710 & $21250\pm340$ & $6.61\pm0.05$ & $18.51$ & $-0.18\pm0.02$ & $0.68\pm0.18$ \\
1800183147814621056\footref{footnote:pelisoli} & 21:51:11.472 & $+$27:30:14.450 & $11900\pm190$ & $5.26\pm0.05$ & $17.34$ & $ 0.02\pm0.01$ & $0.58\pm0.08$ \\
1769840196004315008\footref{footnote:pelisoli} & 21:57:15.929 & $+$16:27:24.638 & $16330\pm470$ & $7.91\pm0.08$ & $18.63$ & $ 0.07\pm0.02$ & $1.85\pm0.21$ \\
1896689409392725504 & 22:01:41.417 & $+$29:30:41.105 & $18680\pm530$ & $7.85\pm0.09$ & $18.93$ & $ 0.55\pm0.02$ & $1.62\pm0.22$ \\
1892667430216516864 & 22:02:04.086 & $+$26:24:47.250 & $18190\pm430$ & $7.33\pm0.07$ & $19.12$ & $-0.11\pm0.02$ & $1.41\pm0.24$ \\
1776485128886007296\footref{footnote:pelisoli} & 22:02:25.343 & $+$18:19:43.417 & $32620\pm3360$ & $5.12\pm0.60$ & $16.30$ & $-0.31\pm0.01$ & $0.69\pm0.06$ \\
2614307800931271552 & 22:07:40.452 & $-$10:03:19.336 & $12100\pm350$ & $7.65\pm0.12$ & $18.41$ & $-0.07\pm0.02$ & $2.02\pm0.26$ \\
2620773547777880832 & 22:09:24.045 & $-$05:47:08.203 & $24640\pm390$ & $7.26\pm0.05$ & $18.31$ & $-0.21\pm0.02$ & $1.44\pm0.16$ \\
6511369569779620224 & 22:14:39.022 & $-$51:47:41.935 & $31110\pm510$ & $8.07\pm0.07$ & $17.89$ & $-0.37\pm0.01$ & $2.21\pm0.10$ \\
6567075776645652352 & 22:20:02.473 & $-$44:42:53.798 & $24380\pm540$ & $7.53\pm0.08$ & $18.56$ & $-0.21\pm0.02$ & $1.74\pm0.17$ \\
2703680916400302208 & 22:21:23.987 & $+$02:38:36.859 & $21430\pm660$ & $7.49\pm0.10$ & $18.63$ & $-0.19\pm0.04$ & $1.42\pm0.24$ \\
2679330242833214848 & 22:21:30.821 & $+$01:20:55.406 & $30150\pm550$ & $7.89\pm0.08$ & $16.88$ & $-0.30\pm0.01$ & $3.70\pm0.09$ \\
2219113776832300416\footref{footnote:tess} & 22:21:40.399 & $+$67:14:47.364 & $11980\pm320$ & $7.42\pm0.14$ & $18.47$ & $ 0.12\pm0.01$ & $2.72\pm0.12$ \\
2626677723354947968 & 22:21:44.381 & $-$04:08:59.770 & $16320\pm2310$ & $5.79\pm0.39$ & $16.38$ & $ 0.24\pm0.01$ & $16.51\pm0.07$ \\
6822591421263379200\footref{footnote:pelisoli} & 22:22:21.437 & $-$18:08:57.797 & $23380\pm550$ & $7.48\pm0.07$ & $18.07$ & $-0.28\pm0.02$ & $1.52\pm0.18$ \\
2737084320170352256 & 22:24:03.905 & $+$16:04:04.030 & $19830\pm520$ & $7.78\pm0.08$ & $18.91$ & $-0.19\pm0.03$ & $1.75\pm0.28$ \\
1881612734154248320 & 22:24:46.707 & $+$27:16:44.886 & $ 9400\pm160$ & $8.39\pm0.19$ & $16.54$ & $ 0.37\pm0.01$ & $14.57\pm0.10$ \\
2737737292637935232\footref{footnote:tess} & 22:29:35.774 & $+$17:45:47.156 & $11780\pm240$ & $8.08\pm0.17$ & $16.63$ & $-0.04\pm0.01$ & $11.85\pm0.06$ \\
2609161983433881088 & 22:35:47.347 & $-$09:09:50.645 & $21650\pm600$ & $7.38\pm0.08$ & $19.37$ & $-0.27\pm0.04$ & $1.56\pm0.33$ \\
2656512108086228096 & 22:45:08.146 & $+$03:01:11.093 & $17580\pm440$ & $7.69\pm0.08$ & $18.90$ & $ 0.24\pm0.03$ & $2.53\pm0.29$ \\
1886328505164612864\footref{footnote:pelisoli} & 22:57:02.142 & $+$30:23:38.497 & $16600\pm260$ & $6.94\pm0.05$ & $18.54$ & $ 0.17\pm0.01$ & $2.50\pm0.17$ \\
2663625089324446976 & 23:01:09.045 & $+$05:46:12.148 & $16610\pm380$ & $7.75\pm0.07$ & $18.95$ & $-0.13\pm0.03$ & $1.61\pm0.29$ \\
2411122140926644736 & 23:03:45.818 & $-$14:14:47.072 & $39520\pm3620$ & $7.68\pm0.39$ & $16.96$ & $-0.02\pm0.01$ & $2.48\pm0.08$ \\
2663797162896215680 & 23:03:58.370 & $+$06:36:08.298 & $13050\pm570$ & $7.71\pm0.10$ & $18.88$ & $ 0.17\pm0.02$ & $2.04\pm0.22$ \\
1931775516926570112\footref{footnote:pelisoli} & 23:05:02.358 & $+$43:02:45.524 & $33020\pm1980$ & $6.29\pm0.38$ & $16.26$ & $-0.200\pm0.004$ & $0.60\pm0.07$ \\
2658636742508253568\footref{footnote:pelisoli} & 23:06:37.878 & $+$02:24:29.606 & $ 9720\pm140$ & $5.42\pm0.08$ & $17.19$ & $ 0.08\pm0.01$ & $1.04\pm0.10$ \\
6378433631820188928 & 23:16:31.886 & $-$75:01:46.337 & $28470\pm2360$ & $7.75\pm0.45$ & $17.71$ & $-0.15\pm0.01$ & $2.29\pm0.08$ \\
2633584713667253376 & 23:18:56.489 & $-$05:08:25.102 & $21480\pm340$ & $7.82\pm0.05$ & $18.47$ & $-0.18\pm0.01$ & $1.73\pm0.25$ \\
6501481249394275840 & 23:20:18.897 & $-$51:45:36.720 & $15880\pm690$ & $7.42\pm0.15$ & $18.85$ & $-0.17\pm0.02$ & $1.99\pm0.19$ \\
2825856243296564608 & 23:22:08.733 & $+$21:03:52.812 & $16680\pm250$ & $6.76\pm0.04$ & $19.09$ & $ 0.07\pm0.02$ & $0.83\pm0.25$ \\
2660773437198409216 & 23:22:30.203 & $+$05:09:42.059 & $18520\pm330$ & $7.18\pm0.05$ & $18.75$ & $-0.18\pm0.02$ & $1.16\pm0.22$ \\
6528026659841832576\footref{footnote:tess} & 23:31:55.589 & $-$47:17:56.234 & $24410\pm2320$ & $7.40\pm0.33$ & $16.89$ & $-0.00\pm0.01$ & $3.20\pm0.06$ \\
6523373973310305664\footref{footnote:tess} & 23:39:23.812 & $-$48:53:50.168 & $30870\pm690$ & $7.58\pm0.12$ & $17.00$ & $-0.33\pm0.01$ & $2.20\pm0.06$ \\
2387618774212898944 & 23:46:11.487 & $-$22:26:28.849 & $17220\pm960$ & $7.87\pm0.18$ & $19.90$ & $ 0.21\pm0.05$ & $1.76\pm0.46$ \\
2639862585760032896 & 23:47:32.871 & $-$02:24:33.894 & $16250\pm270$ & $7.05\pm0.05$ & $19.03$ & $-0.10\pm0.02$ & $1.35\pm0.27$ \\
2867393662290131840\footref{footnote:pelisoli} & 23:51:17.370 & $+$29:34:35.573 & $22310\pm730$ & $7.46\pm0.11$ & $18.07$ & $-0.21\pm0.01$ & $2.18\pm0.14$ \\
6379389141784320640\footref{footnote:pelisoli} & 23:57:36.404 & $-$74:48:00.947 & $22440\pm420$ & $7.40\pm0.06$ & $18.14$ & $-0.19\pm0.01$ & $1.76\pm0.10$ \\
6386812017799240064\footref{footnote:pelisoli} & 23:58:54.268 & $-$69:56:13.639 & $36100\pm750$ & $7.58\pm0.11$ & $18.20$ & $-0.34\pm0.01$ & $1.06\pm0.12$ \\
2312944342501257472\footref{footnote:pelisoli} & 23:59:12.953 & $-$34:04:00.476 & $24390\pm520$ & $7.87\pm0.07$ & $17.81$ & $-0.31\pm0.01$ & $2.19\pm0.12$ \\
\enddata
\end{deluxetable}

\end{document}